\documentclass[article]{aa} 

\usepackage{graphicx}
\usepackage{natbib}
\bibpunct{(}{)}{;}{a}{}{,} %

\usepackage{amsmath} \usepackage{amssymb} \usepackage{wasysym}

\usepackage{setspace}

\title{Flux and polarization signals of \\
       spatially inhomogeneous gaseous exoplanets}

\author{T. Karalidi\inst{1,2}\thanks{Present address: Steward Observatory, 933 N Cherry Ave, Tucson, AZ 85721, USA} 
   \and D. M. Stam\inst{3}
   \and D. Guirado\inst{1}}
        
\institute{
  SRON - Netherlands Institute for Space Research,
  Sorbonnelaan 2, 3584 CA, Utrecht, the Netherlands 
  \and
  Leiden Observatory, Leiden University, 2300 RA, 
  Leiden, the Netherlands 
  \and
  Faculty of Aerospace Engineering, Technical University Delft, 
  Kluyverweg 1, 2629 HS, Delft, the Netherlands  
  }

\date{Received 18 March 2013 / Accepted 22 May 2013}
\begin{document}


\abstract{}{We present numerically calculated, disk--integrated, 
  spectropolarimetric signals of starlight that is reflected by
  vertically and horizontally inhomogeneous gaseous exoplanets.
  We include various spatial features that are present
  on Solar System's gaseous planets: belts and zones, cyclonic spots,
  and polar hazes, to test whether such features leave traces
  in the disk--integrated flux and polarization signals.}
  {Broadband flux and polarization signals of starlight that is reflected
  by gaseous exoplanets are calculated
  using an efficient, adding--doubling radiative transfer code, that 
  fully includes single and multiple scattering and polarization. 
  The planetary model atmospheres are vertically inhomogeneous 
  and can be horizontally inhomogeneous, and
  contain gas molecules and/or cloud and/or aerosol particles.}
  {The broadband flux and polarization signals
  are sensitive to cloud top pressures, although
  in the presence of local pressure differences, such as in belts and clouds,
  the flux and polarization phase functions have similar shapes as
  those of horizontally homogeneous planets. 
  Fitting flux phase functions of a planet with belts and zones using 
  a horizontally homogeneous planet could theoretically yield 
  cloud top pressures that differ by a few hundred mbar from those 
  derived from fitting polarization phase functions. In practise, 
  however, observational errors and uncertainties in cloud properties
  would make such a fit unreliable.
  A cyclonic spot like Jupiter's Great Red Spot, covering a few
  percent of the disk, located 
  in equatorial regions, and rotating in and out of the observer's
  view yields a temporal variation of a few percent in the broadband flux 
  and a few percent in the degree of polarization.
  Polar hazes leave strong traces in the polarization of
  reflected starlight in spatially resolved observations, especially
  seen at phase angles near 90$^\circ$. Integrated across the planetary disk, 
  polar hazes that cover only part of the planetary disk, change the 
  broadband degree of polarization of the reflected light by a few percent. 
  Such hazes have only small effects on locally reflected broadband fluxes
  and negligible effects on disk--integrated broadband fluxes.}
  {Deriving the presence of belts and zones in the atmospheres of gaseous 
  exoplanets from broadband flux and polarization observations will be
  extremely difficult.
  Cyclonic spots could leave temporal changes 
  in the broadband flux and polarization signals of a few percent.
  Polar hazes that cover a fraction of the planetary disk, and that are
  composed of small, Rayleigh scattering particles, change the broadband
  degree of polarization by at most a few percent.}

\keywords{exoplanets - polarization }

\titlerunning{Signals of spatially inhomogeneous gaseous exoplanets}
\authorrunning{T. Karalidi et al.}

\maketitle


\section{Introduction}

Since the discovery of the first exoplanet orbiting a main sequence
star by \citet{mayorqueloz95}, more than 850 exoplanets have been
detected up to today.
The refinement of the detection methods and the instrumentation, 
such as the highly successful space missions CoRoT (COnvection,
ROtation \& planetary Transits) \citep{baglin06,magali10} and Kepler
\citep{koch98}, and ground-based telescope instruments like
HARPS (High Accuracy Radial Velocity Planet Searcher) \citep{pepe04}
have led to an almost exponential increase of the number of detected
planets per year. 

The next step of exoplanet research is the characterization of detected
exoplanets: what is the composition and structure of their atmospheres,
and, for rocky exoplanets, their surface?
In the near future, instruments like SPHERE
(Spectro--Polarimetric High--Contrast Exoplanet Research) 
\citep{dohlen08,roelfsema11}
on the VLT (Very Large Telescope) and GPI (Gemini Planet Imager)
\citep{macintosh08} on the Gemini North telescope, and more in the
future, EPICS (Exoplanet Imaging Camera and Spectrograph) 
\citep[][]{kasper10} on the E-ELT,
will characterize gaseous exoplanets in relatively wide orbits around
their stars, and possibly super-Earths around nearby stars, using
combinations of spectroscopy and broadband polarimetry.

In exoplanet research, polarimetry helps to detect a planet because the
direct stellar light is usually unpolarized \citep[see][]{kemp87}, 
while the starlight that has been reflected by a planet will usually
be (linearly) polarized because it has been scattered by atmospheric
particles and/or it has been reflected by the planetary surface (if 
present). Polarimetry helps to confirm the detection of an exoplanet,
because background sources will usually be unpolarized, or have a
direction of polarization that excludes a relation with the star.
A first detection of the polarization signal of an exoplanet was
claimed by \citet{berdyugina08, berdyugina11}.
The power of polarization in characterizing planetary atmospheres and
surfaces has been demonstrated through observations of Solar System planets 
(including Earth itself)\citep[see, for example][]{hansenhovenier74, hansentravis74,
mishchenko90, tomasko09}. For gaseous and terrestrial--type exoplanets
numerical calculations have clearly shown the added information 
on planetary characteristics that can be derived from polarimetry
\citep[e.~g.~][]{stam03,stamhovenier04, saar03, seager00, stam08, karalidi11}.

The numerical studies mentioned above pertain to planetary model atmospheres
that are vertically inhomogeneous, but horizontally homogeneous. Partly, this
has to do with the computational effort: fully including all orders of scattering
and polarization in radiative transfer calculations requires orders more 
computing time than the scalar radiative transfer calculations that are
commonly used to model reflected fluxes (note that ignoring polarization
in radiative transfer calculations introduces errors in calculated fluxes, see e.g. 
\citet[][]{stam05} and references therein).

In this paper, we present numerically calculated total flux and 
polarization signals of unpolarized incident starlight that is reflected by
gaseous exoplanets that are both vertically and horizontally inhomogeneous.
For the horizontal inhomogeneities, we use banded structures similar to
the belts and zones that circle Jupiter and Saturn, cyclonic spots such
as the long lived Great Red Spot on Jupiter and the Great Dark Spot 
that Voyager--2 observed on Neptune, and polar hazes such as those
covering the north and south poles of Jupiter and Saturn.
In particular the latter hazes are known to strongly polarize the incident sunlight,
as observed at small phase angles from Earth \citep[][]{schmid11},
and at intermediate phase angles from spacecraft observations
\citep[see e.g.][]{west83, smith84}. 

For exoplanets, observations will yield disk--integrated flux and
polarization signals. As we will explore in this paper, local 
horizontal inhomogeneities might reveal
themselves through temporal variations in flux and/or polarization 
signals when planets rotate with respect to the observer, or because 
the flux and/or polarization signals of the horizontally inhomogeneous
planets deviate from those of horizontally homogeneous planets.
The dependence of a total flux signal on the composition and
structure of a planetary atmosphere is different than that of 
a polarization signal. Therefore, a horizontally inhomogeneous 
planet's total flux and polarization phase functions could be
similar to those of different horizontally homogeneous planets.
Such differences could be used to detect spatial features like
belts and zones, cyclonic spots, and/or polar hazes.

This paper is organized as follows. 
In Sect.~\ref{section2}, we describe polarized light, 
our radiative transfer algorithm, and the model planetary atmospheres.
In Sect.~\ref{sec:ss_props}, we present the single
scattering properties of the cloud and haze particles in the model
atmospheres.
Section~\ref{sec:gaseousplanets} shows the calculated flux
and polarization signals of different types of spatially inhomogeneous
model planets: with zones and belts (Sect.~\ref{sec:zonesbelts}), 
cyclonic spots (Sect.~\ref{sec:cyclonicspots}), and polar hazes
(Sect.~\ref{sec:polarhazes}).
Section~\ref{sect_summary}, finally, contains a summary and 
our conclusions.


\section{Description of the numerical simulations}
\label{section2}

\subsection{Definitions of flux and polarization}

Starlight that has been reflected by a planet can be 
described by a flux vector $\pi\vec{F}$, as follows
\begin{equation}
\label{eq:first}
   \pi\vec{F}= \pi\left[\begin{array}{c} 
               F \\ Q \\ U \\ V \end{array}\right],
\end{equation}
where parameter $\pi F$ is the total flux, parameters $\pi Q$ and $\pi
U$ describe the linearly polarized flux and parameter $\pi V$ the
circularly polarized flux \citep[see e.g.][]{hansentravis74,hovenier04}. 
Although not explicitly shown in Eq.~\ref{eq:first}, 
all four parameters depend on the wavelength $\lambda$, 
and their dimensions are W~m$^{-2}$m$^{-1}$. Parameters
$\pi Q$ and $\pi U$ are defined with respect to a reference plane, for
which we chose the planetary scattering plane, i.e. the plane
through the centers of the star, the planet and the observer.
Parameter $\pi V$ of starlight that is reflected by a planet is generally
small \citep[][]{hansentravis74} and we will ignore it in our
simulations. This can be done without introducing significant errors in
our calculated fluxes $\pi F$, $\pi Q$ and $\pi U$ \citep{stam05}.
The degree of (linear) polarization $P$ of flux vector $\pi \vec{F}$
is defined as follows
\begin{equation}
   P=\frac{\sqrt{Q^{2}+U^{2}}}{F},
\label{eq:poldef}
\end{equation}
which is independent of the choice of reference plane.

Unless stated otherwise, we assume that the starlight that is incident on
a model planet is unpolarized \citep[][]{kemp87} and that the model 
planets are mirror-symmetric with respect
to the planetary scattering plane. In that case, $\pi U$
equals zero, and we can use an alternative definition for the degree
of polarization, namely
\begin{equation}
   P_\mathrm{s}= - \frac{Q}{F},
\label{eq:polsign}
\end{equation}
with the subscript {\em s} referring to {\em signed}. 
For $P_\mathrm{s} > 0$ ($< 0$), the reflected light is
polarized perpendicular (parallel) to the reference plane.

We will present calculated fluxes that are normalized such that at a
planetary phase angle $\alpha$ equal to 0$^\circ$ (i.e. seen from the
middle of the planet, the angle between the star and the observer
equals 0$^\circ$), the
total reflected flux $\pi F$ equals the planet's geometric albedo
$A_\mathrm{G}$ \citep[see e.g.][]{stamhovenier04}. We will indicate the hence
normalized total flux by $\pi F_\mathrm{n}$ and the associated 
linearly polarized fluxes by $\pi Q_\mathrm{n}$ and $\pi U_\mathrm{n}$. 
The normalized fluxes that we present in this paper can straightforwardly be
scaled to absolute fluxes of a particular planetary system by multiplying 
them with $r^2/d^2$, where $r$ is the spherical planet's radius and $d$ the
distance between the planet and the observer, and with the
stellar flux that is incident on the planet. In our calculations, we
furthermore assume that the distance between the star and the planet
is large enough to assume that the incident starlight is
uni-directional. Since the degree of polarization $P$ (or $P_\mathrm{s}$) 
is a relative measure, it doesn't require any scaling.

Our calculations cover phase angles $\alpha$ from 0$^\circ$ to 
180$^\circ$. Of course, the range of phase angles an exoplanet
exhibits as it orbits its star, depends on the orbital 
inclination angle. Given an orbital inclination angle $i$ (in degrees), 
one can observe the exoplanet at phase angles ranging from 
$90^\circ - i$ to $90^\circ + i$. Thus, an exoplanet in a face--on orbit
($i=0^\circ$) would always be observed at a phase angle equal to 90$^\circ$,
while the phase angles of an exoplanet in an edge--on orbit 
($i=90^\circ$) range from 0$^\circ$ to 180$^\circ$, the complete range
that is shown in this paper.
Note that the actual range of phase angles an exoplanet can be
observed at will depend strongly on the observational technique 
that is used, and e.g. on the angular distance between a star and its
planet.

\subsection{Our radiative transfer code}

Our radiative transfer code to calculate the total and polarized
fluxes that are reflected by model planets is based on the efficient
adding--doubling algorithm described by \citet{dehaan87}. It fully
includes single and multiple scattering and polarization, and assumes
that locally, the planetary atmosphere is plane--parallel. 
We will use a version of the code that applies to horizontally
homogeneous planets \citep{stamhovenier04,stam06,stam08}, 
and a (more computing-time-consuming) version that applies to horizontally
inhomogeneous planets \citep{karalidi12b}. 
In the latter code, a model planet is divided into pixels that are 
small enough to be considered horizontally homogeneous. 
Reflected stellar fluxes are then calculated for all pixels that are both 
illuminated and visible to the observer and then summed up to acquire 
the disk--integrated total and polarized reflected fluxes. 
Since the adding--doubling code uses the local meridian plane 
(which contains both the local zenith direction and the 
propagation direction of the reflected light) as the reference plane, 
we have to rotate locally calculated flux vectors to the planetary
scattering plane before summing them up. Following
\citet{karalidi12b}, we divide our model planets into pixels of
$2^\circ\times2^\circ$ (latitude $\times$ longitude).

\subsection{Our model planets}
\label{sect:modelplanets}

The atmospheres of our model planets consist of homogeneous,
plane--parallel layers that contain gases and, optionally, clouds or
hazes. Here, we use the term 'haze' for optically thin
layers of submicron--sized particles, while 'clouds' are thicker and composed
of larger particles. The model atmospheres are bounded below by black
surfaces, i.e. no light is entering the atmospheres from below. 
The ambient atmospheric temperature and pressure profiles are representative 
for midlatitudes on Jupiter \citep[see][]{stamhovenier04}. Given the 
temperatures and pressures across an atmospheric layer, 
and the wavelength $\lambda$, the gaseous scattering
optical thickness of each atmospheric layer is calculated according to
\citet[][]{stam99}, using a depolarization factor that is representative
for hydrogen--gas, namely 0.02 \citep[see][]{hansentravis74}. 
At $\lambda=0.55$~$\mu$m, the total
gaseous scattering optical thickness of our model atmosphere is 5.41. 
We ignore absorption by methane, and choose wavelengths in the
continuum for our calculations. In particular, when broad band filters
are being used for the observations, the contribution of reflected
flux from continuum wavelengths will contribute most to the 
measured signal.

The physical properties of the clouds and hazes across a planet like
Jupiter vary in time \citep[for an overview, see e.g.][]{west04}. 
Here, we use a simple atmosphere model that suffices to show the effects 
of clouds and hazes on the flux and polarization signals of Jupiter--like
exoplanets. Our model atmospheres have an optically thick tropospheric
cloud layer that is composed of ammonia ice particles (their properties
are presented in Sect.~\ref{sec:ss_props}). The bottom of this cloud
layer is at a pressure of 1.0~bar. We vary the top of the cloud between
0.1 and 0.5 bars. The cloud top pressure of 0.1 bars is representative for 
the so--called zonal bands on Jupiter. 
In the zones, the clouds typically rise up higher into the atmosphere 
than in the adjacent belts where the cloud top pressures
can be up to a few hundred~mbar higher \citep[see][]{ingersoll04}.
We set the optical thickness of the clouds in the zones at 21 (at 
0.75 $\mu$m) and in the belts it varies from 20 to 6.02 as the cloud 
top pressure varies from 0.1~bar to 0.5~bar.
On Jupiter, the clouds are overlaid by a stratospheric, photochemically
produced haze layer.
The haze layers over in particular both polar regions, provide strong
polarization signals indicating that they consist of small aggregated 
particles \citep[][]{west91}.
To avoid introducing too many variables, we only use haze layers over the 
polar regions of our model planets.

We will present results for horizontally homogeneous model planets and
for model planets with bands of clouds divided into zones and
belts that run parallel to the equator, which lies in the planet's
equatorial plane. Our banded model planets are mirror--symmetric:
measured from the equator in either the northern or the southern
direction, we chose the latitudes that 
bound the belts and zones as follows: 0$^\circ$--8$^\circ$ (zone),
8$^\circ$--24$^\circ$ (belt), 24$^\circ$--40$^\circ$ (zone),
40$^\circ$--60$^\circ$ (belt), 60$^\circ$--90$^\circ$ (zone).
These latitudes correspond roughly to the most prominent cloud bands
of Jupiter \citep[see e.g.][]{depaterlissauer02}.
The northern and southern polar 
hazes extend upward, respectively downward, from a latitude of 60$^\circ$.
Vertically these hazes extend between $\sim$0.0075 bar and $\sim$0.0056 bar, 
and we give them an optical thickness of 0.2 at 0.55~$\mu$m. The single
scattering albedo of the haze particles is 0.995 at 0.55~$\mu$m.


\section{Single scattering properties of the cloud and haze particles}
\label{sec:ss_props}

\subsection{The tropospheric cloud particles}
\label{sec:nh3ice}

Thermodynamic models of the jovian atmosphere indicate that the upper
tropospheric cloud layers should consist of ammonia ice particles
\citep[see for example][]{sato79,simonmiller01,depaterlissauer02}. 
Galileo NIMS and Cassini CIRS data, however, indicated that spectrally 
identifiable ammonia ice clouds cover only very small regions on the 
planet \citep[see, respectively][]{baines02,wong04}. 
As put forward by e.g.\ \citet{atreya05}, this apparent contradiction 
could be explained if the ammonia ice particles are coated by in particlar 
hydrocarbon haze particles settling from the stratosphere.
Thus, only the highest and freshest ammonia ice clouds would
show identifiable spectral features.
\citet{atreya05} also mention that the strength of the spectral
features would depend on the sizes and shapes of the ice crystals.
In this paper, we assume that the upper tropospheric clouds in our 
model atmospheres are indeed composed of ammonia ice particles,
without modelling specific spectral features.


\begin{figure}
\centering
\includegraphics[width=85mm]{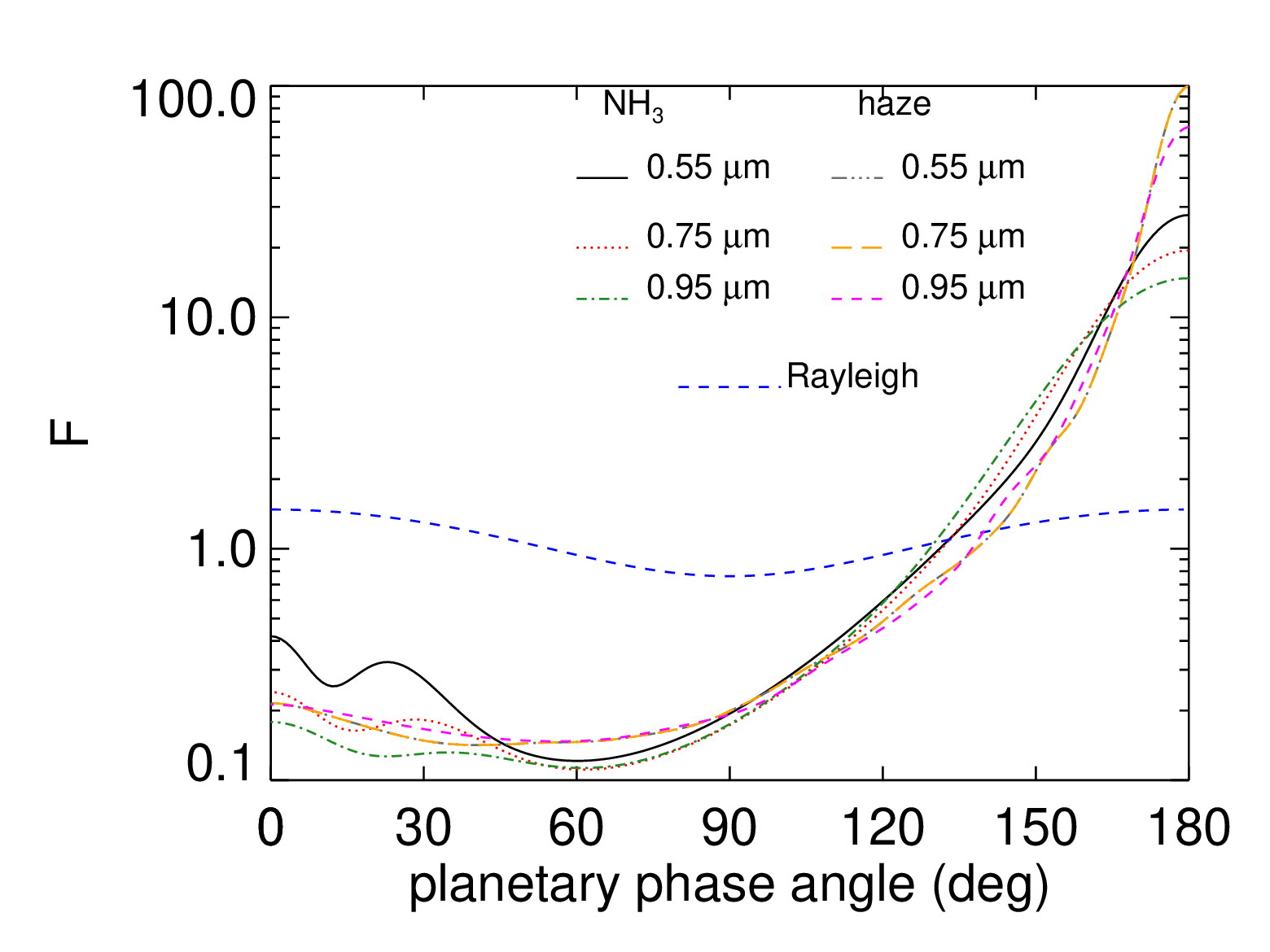}
\hspace{0.8cm}
\centering
\includegraphics[width=85mm]{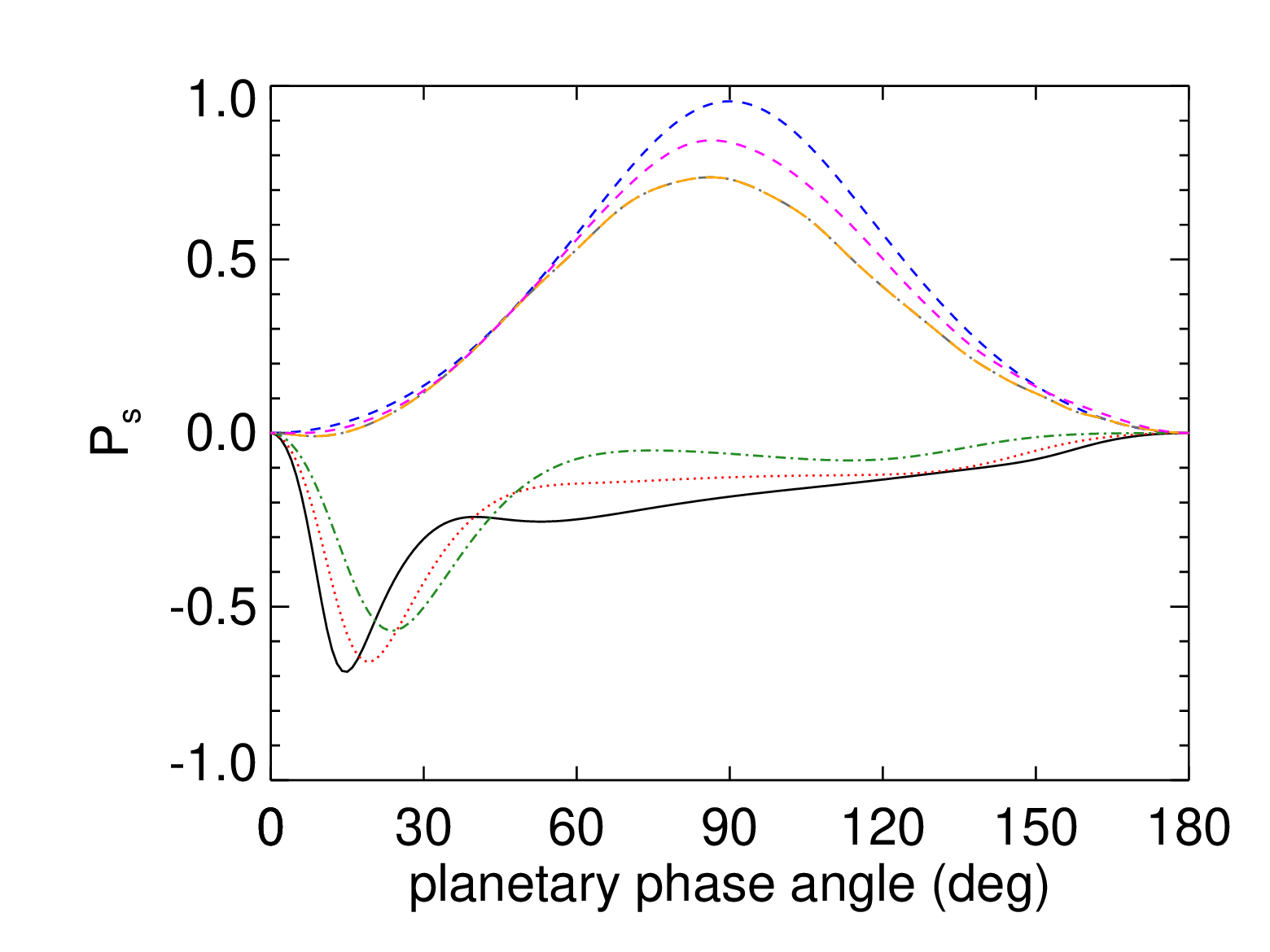}
\caption{Single scattering $F$ and $P_\mathrm{s}$ of our ammonia (NH$_3$) 
  ice cloud particles and the polar haze particles as functions of the 
  planetary phase angle $\alpha$ at three different wavelengths.
  The curves for the NH$_3$ ice particles are: black, solid line 
  ($\lambda=0.55~\mu$m), red, dotted line (0.75~$\mu$m), 
  green, dashed--dotted line (0.95~$\mu$m). 
  The curves for the polar haze particles are:
  grey, dashed--triple--dotted line (0.55~$\mu$m),
  orange, long--dashed line (0.75~$\mu$m),
  magenta, dashed line (0.95~$\mu$m).
  The blue, dashed lines are the Rayleigh scattering curves at 
  0.55~$\mu$m.}
\label{fig:ss_ice}
\end{figure}

Our ammonia ice particles are assumed to be spherical with a
refractive index of $n=1.48~+~0.01i$ (assumed to be constant
across the spectral region of our interest) 
(as adopted from \citet{romanescu10} and \citet{gibson05}, 
for the real and imaginary part respectively)
and with their sizes described by a standard size distribution 
\citep[see][]{hansentravis74}
with and effective radius $r_\mathrm{eff}$ of 0.5$\mu$m, 
and an effective variance $v_\mathrm{eff}$ of 0.1
\citep[][]{stamhovenier04}. We calculate the single scattering
properties of the ammonia ice particles using Mie theory as
described by \citet{derooijvdstap87}.

Figure~\ref{fig:ss_ice} shows the flux and degree of linear
polarization $P_\mathrm{s}$ of unpolarized incident light with
$\lambda$=0.55~$\mu$m, 0.75~$\mu$m, and 0.95~$\mu$m, respectively, 
that is singly scattered by the ice particles as functions of the 
planetary phase angle $\alpha$. Note that $\alpha= 180^\circ - \Theta$, 
with $\Theta$ the conventional single scattering angle, defined
as $\Theta=0^\circ$ for forward scattered light. 
All scattered fluxes have been normalized such that their
average over all scattering directions equals one 
\citep[see Eq.~2.5 of][]{hansentravis74}.
At $\alpha=0^\circ$ (180$^\circ$) the light is
scattered in the backward (forward) direction. For comparison, we
have added the curves for (Rayleigh) scattering by gas molecules
at $\lambda=0.55~\mu$m (these curves are fairly wavelength independent).
As can be seen in the figure, our spherical ice particles are moderately 
forward scattering and the scattered fluxes 
show a prominent feature (a local minimum) around 
$\alpha=12^\circ$ at $\lambda=0.55~\mu$m, around $20^\circ$
at 0.75~$\mu$m, and (much less pronounced)
around $25^\circ$ at 0.95~$\mu$m.

The degree of polarization $P_\mathrm{s}$ of the light that is singly
scattered by our ammonia ice particles is negative across almost the
whole phase angle range. The light is thus polarized parallel to the
scattering plane, which contains both the incident and the scattered
beams. The local minima in scattered fluxes have associated  
local minima in $P_\mathrm{s}$ (at slightly shifted values of $\alpha$).

We should note here that the vast majority of giant planets 
imaged so far, are too hot to contain ammonia ice particles 
in their atmospheres \citep[see e.g.][]{bonnefoy13, konopacky13}. 
These planets though, are so far away from their parent star and 
so hot that the contribution of the reflected starlight to 
the disk--integrated signal is very small in comparison 
to the thermal radiation of the planets. Even though 
the thermal radiation can also be polarized 
(through scattering by the cloud particles, see \citet[][]{dekok11}) 
these planets are out of the 
scope of this paper.

\subsection{Polar haze particles} 
\label{sec:hazeparts}

\begin{figure}
\centering
\includegraphics[width=85mm]{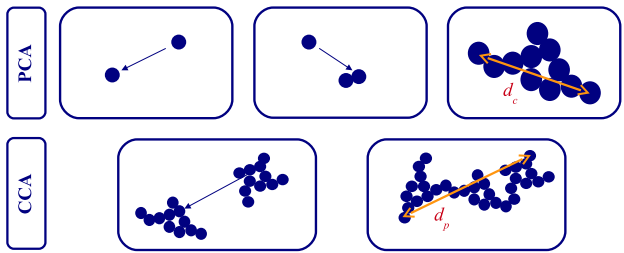}
\caption{Sketch of the two aggregation--mechanisms to build the polar
  haze particles. With PCA,
  identical monomers are sticked together until the maximum distance between 
  two monomers is larger than a given limiting distance $d_\mathrm{c}$. 
  With CCA, several PCA--particles
  are combined until the maximum distance between two monomers of
  the whole particle is larger than $d_\mathrm{p}$.}
\label{fig:particles1}
\end{figure}

We model Jupiter's polar haze particles as randomly oriented aggregates
of equally sized spheres. To generate the aggregates (needed for calculating
the single scattering properties of these particles), we use a
cluster--cluster aggregation (CCA) method that starts with the
formation of particle--cluster aggregates (PCA) by sequentially adding
spheres from random directions to an existing cluster, as shown in the
upper part of Fig.~\ref{fig:particles1}. Next, we combine several
PCA--particles, as shown in the lower part of Fig.~\ref{fig:particles1}. 
For both PCA and CCA, the coagulation process finishes when the maximum
distance between any pair of monomers of the aggregate exceeds a
certain limit (in Fig.~\ref{fig:particles1}: $d_\mathrm{c}$ for PCA
and $d_\mathrm{p}$ for CCA). With the later assumption, we limit the size
of the generated particles, which is needed due to the computational 
limitations of the numerical methods to calculate the single scattering
properties of the particles (see below).
We use CCA--particles rather than PCA--particles in the model atmospheres
because they can yield the high polarization values
that have been observed at the poles of Jupiter \citep[see e.g.][]{schmid11}.
Because PCA--particles are more compact, light is on average scattered 
more within each particle, which decreases the degree of polarization 
of the scattered light.


\begin{figure}
\centering
\includegraphics[width=85mm]{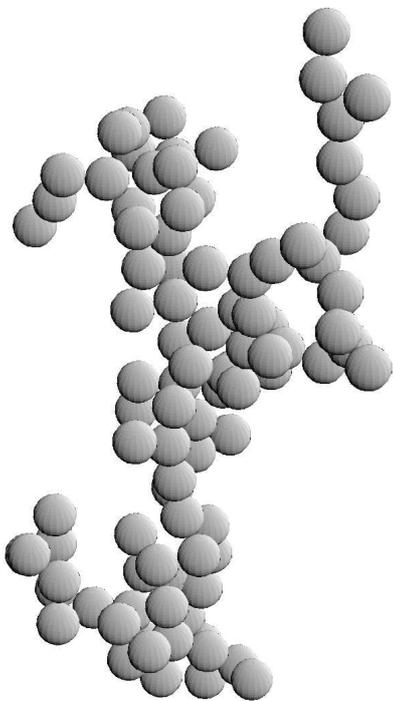}
\caption{A picture of a model aggregate haze particle that we
         used for our single light scattering calculations.}
\label{fig:particles2}
\end{figure}

Figure~\ref{fig:particles2} shows a model aggregate haze particle that 
we generated and for which
we calculated the single scattering matrices and other optical properties.
The particle consists of 94 spherical monomers. The radius of each
monomer is approximately 0.035~$\mu$m, and the 
volume-equivalent-sphere radius of the whole particle 0.16~$\mu$m.
Calculations of the single scattering matrix and other optical properties
of these particles were performed using the T-matrix
theory combined with the superposition theorem \citep[][]{mackowski11},
at $\lambda=0.55$, 0.75 and 0.95~$\mu$m, 
and adopting a refractive index of $1.5+i0.001$, which corresponds to 
that of benzene \citep[suggested to exist on the jovian poles by e.g.][]{friedson02}.  
In Fig.~\ref{fig:ss_ice} we show the flux and polarization of
unpolarized incident light that is singly scattered by the haze 
particles at the three different wavelengths, together with the
Rayleigh curves at $\lambda=0.55~\mu$m.
The single scattering albedo of the particles is 0.995 at 0.55~$\mu$m.

From comparing the different lines in Fig.~\ref{fig:ss_ice},
it is clear that the haze particles are more forward 
scattering than the ammonia ice particles, and that their scattered
flux shows less angular features. 
The degree of linear polarization of the light scattered by the haze
particles is very different from that of the cloud particles:
it is positive at almost all phase angles (hence the light is polarized 
perpendicular to the scattering plane) and it reaches values 
larger than 0.7 near $\alpha=90^\circ$.
The main reason that the polarization phase function of the
haze particles differs strongly from that of the cloud particles
while their flux phase functions are quite similar, is that
the latter depends mostly on the size of the whole particle, 
while the polarization phase function depends more on the
size of the smallest scattering particles, which have radii of
about 0.035~$\mu$m, in the case of the aggregate particles.

The maximum single scattering polarization of our aggregate particles is 
slightly higher than that derived by \citet[][]{west91}. 
This is most likely due to the properties of our haze particles: 
our monomers are smaller than those used by
\citet[][]{west91}, which have radii near 0.06~$\mu$m, sometimes
mixed with monomers with radii of 0.03~$\mu$m.
In addition, the particles in \citet[][]{west91} were generated using the
diffusion-limited aggregation (DLA) method, in which monomers
follow random paths towards the aggregate, and which yields more 
compact particles than those produced by our CCA--method 
\citep[see][]{meakin83}.


\section{Reflected flux and polarization signals of the model planets} 
\label{sec:gaseousplanets}

In this section, we present fluxes and degrees of linear polarization
for three different types of spatial inhomogeneities that occur
on gaseous planets in the Solar System: 
zones and belts (Sect.~\ref{sec:zonesbelts}), 
cyclonic spots (Sect.~\ref{sec:cyclonicspots}), and
polar hazes (Sect.~\ref{sec:polarhazes}). 
We will compare the flux and polarization signals of the spatially
inhomogeneous planets with those of horizontally homogeneous planets
to investigate whether or not such spatial inhomogeneities would be 
detectable.

\subsection{Zones and belts}
\label{sec:zonesbelts}

\begin{figure}
\centering
\includegraphics[width=85mm]{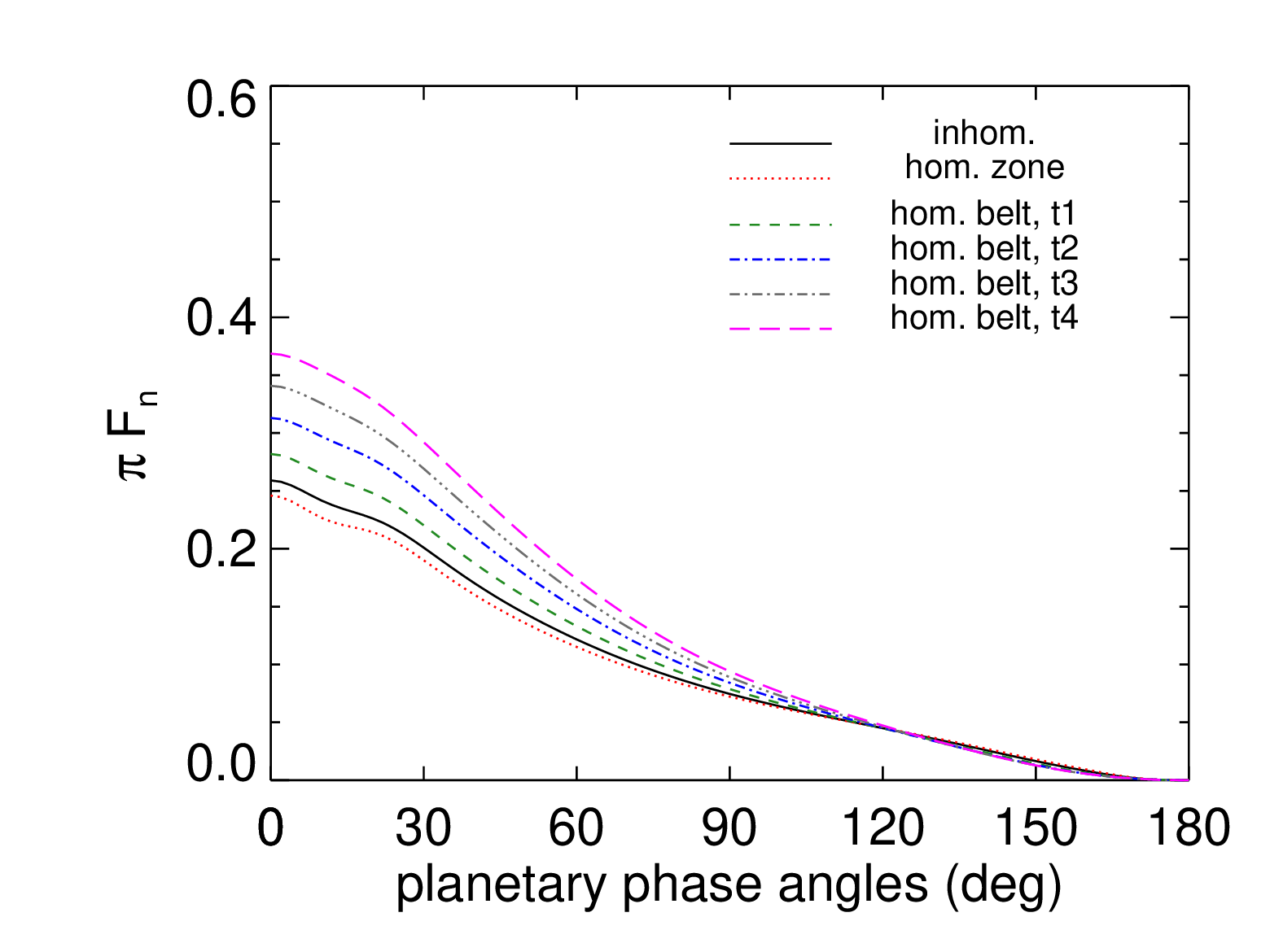}
\hspace{0.8cm}
\centering
\includegraphics[width=85mm]{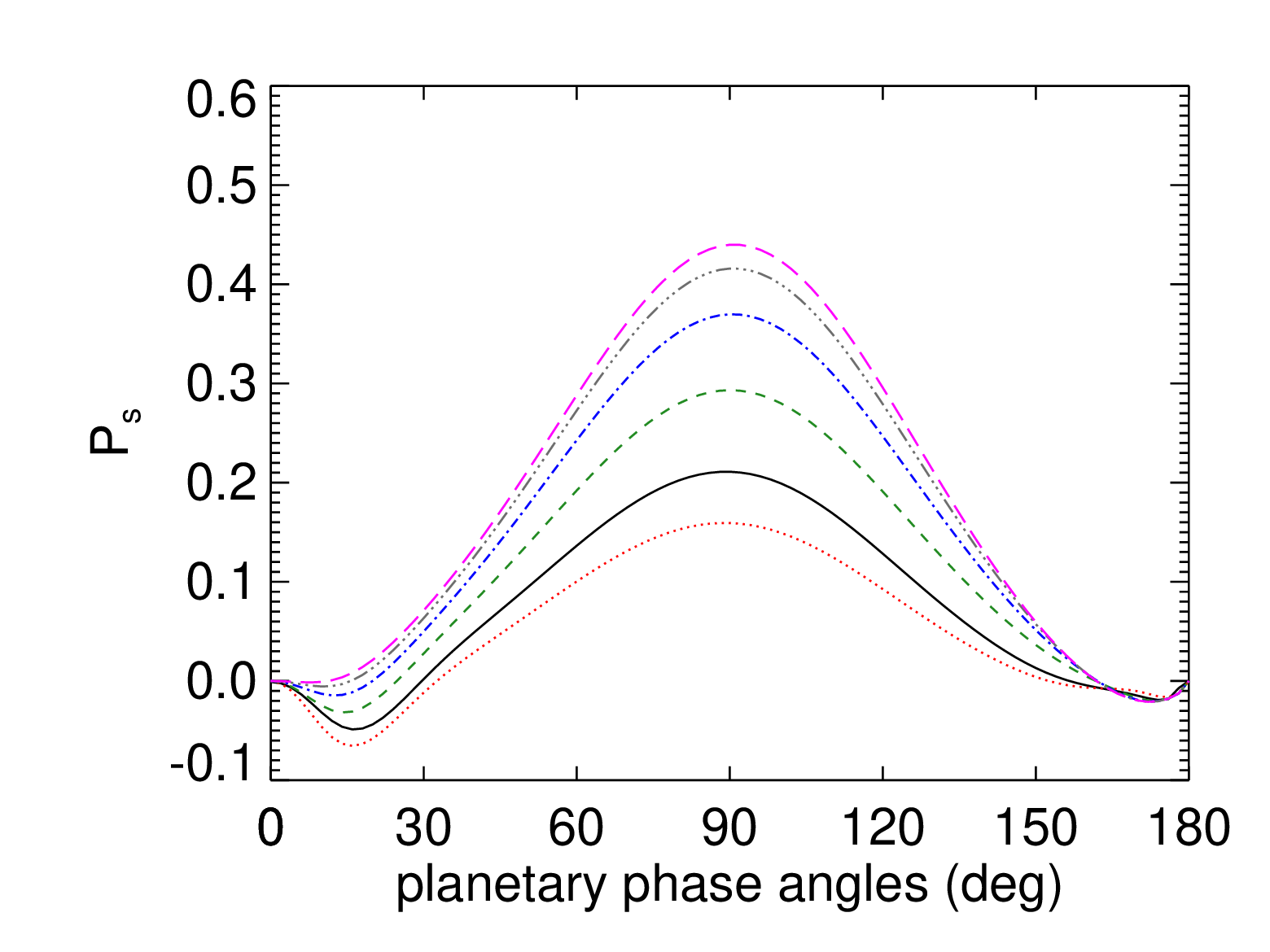}
\caption{Total flux $\pi F_\mathrm{n}$ and degree of polarization 
  $P_\mathrm{s}$ of starlight with $\lambda=0.55$~$\mu$m
  that is reflected by a cloud covered model planet. 
  The bottom of the clouds is at 1.0~bar, while the cloud top pressure
  varies. For the horizontally homogeneous planets, the cloud top
  pressures are as follows:
  0.1~bar (red, dotted line), 0.2~bar (green, dashed line), 
  0.3~bar (blue, dashed--dotted line), 0.4~bar (grey, dashed--triple-dotted line), 
  0.5~bar (magenta, long--dashed line).
  The spatially inhomogeneous planet (black, solid line) has a cloud 
  top pressure of 0.1~bar in the zones, and 0.2~bar in the belts.}
\label{fig:top02_550nm}
\end{figure}

\begin{figure}
\centering
\includegraphics[width=85mm]{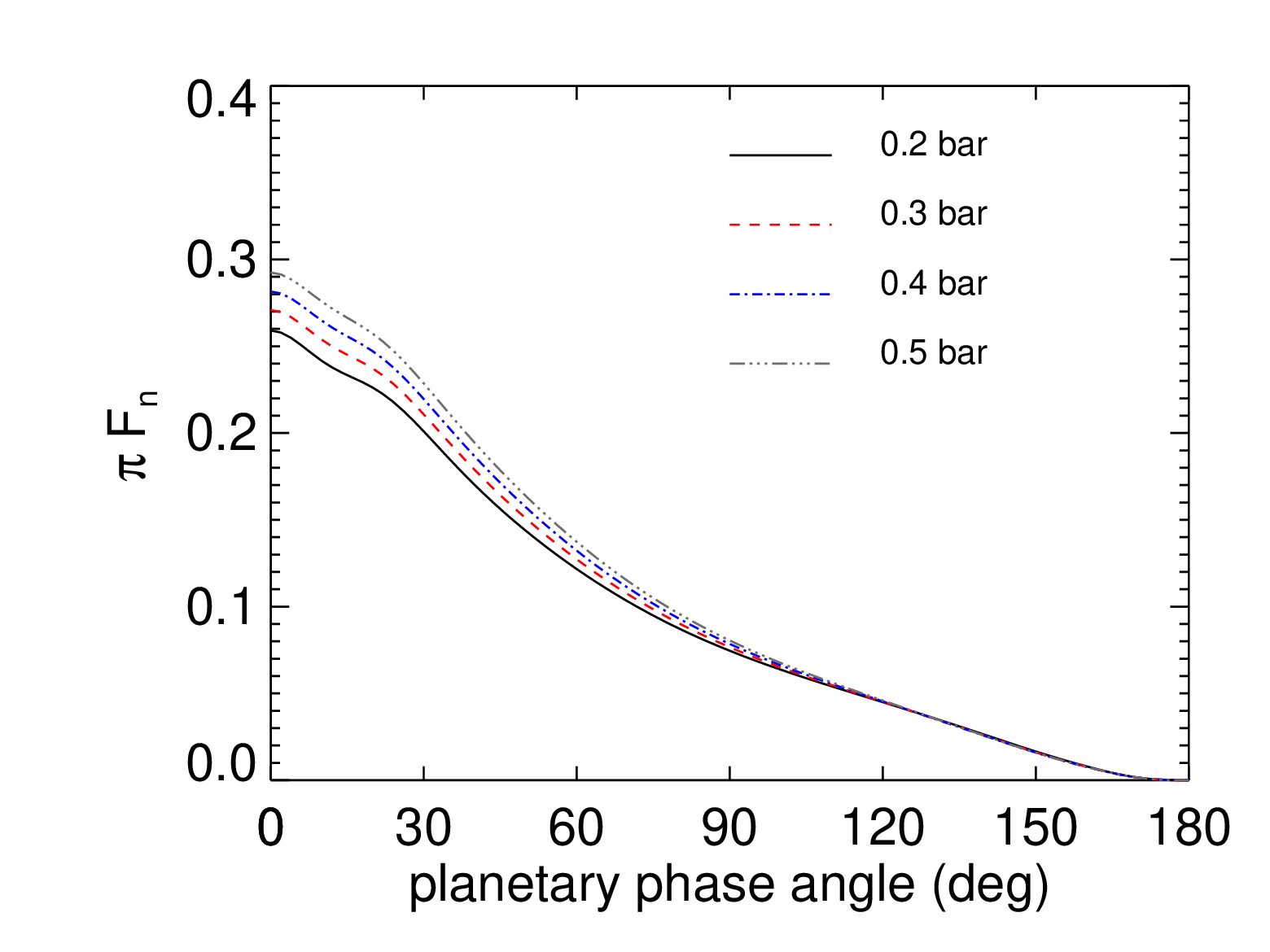}
\hspace{0.8cm}
\centering
\includegraphics[width=85mm]{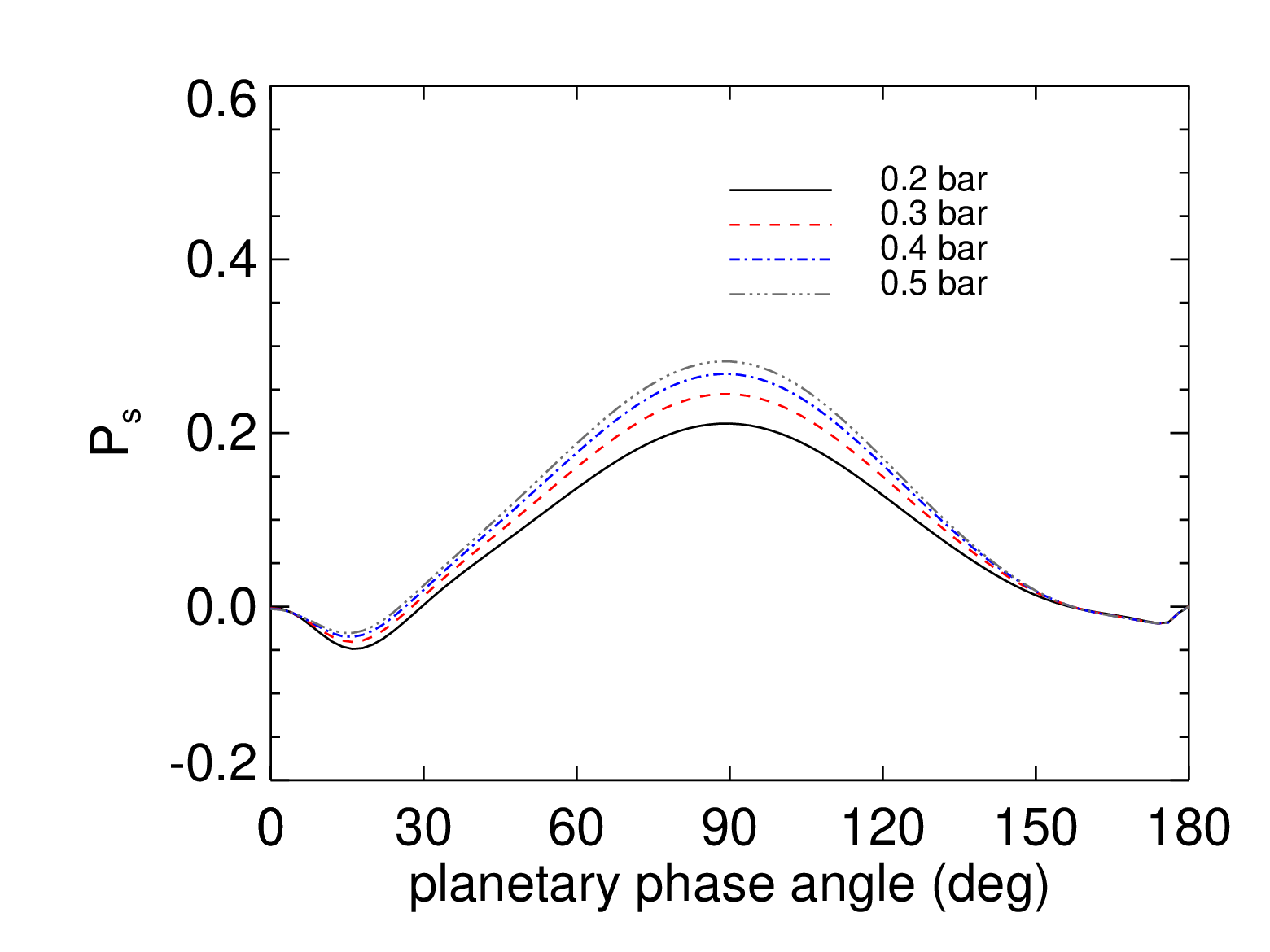}
\caption{Total flux $\pi F_\mathrm{n}$ and degree of polarization 
  $P_\mathrm{s}$ of starlight with $\lambda=0.55$~$\mu$m
  that is reflected by a spatially inhomogeneous model planet. The cloud 
  top pressure is 0.1~bar in the zones and 0.2~bar (black, solid line), 
  0.3~bar (red, dashed line), 0.4~bar (blue, dashed--dotted line) or 0.5~bar (grey, 
  dashed--triple--dotted line)  in the belts. }
\label{fig:top03_550nm}
\end{figure}

\begin{figure}
\centering
\includegraphics[width=85mm]{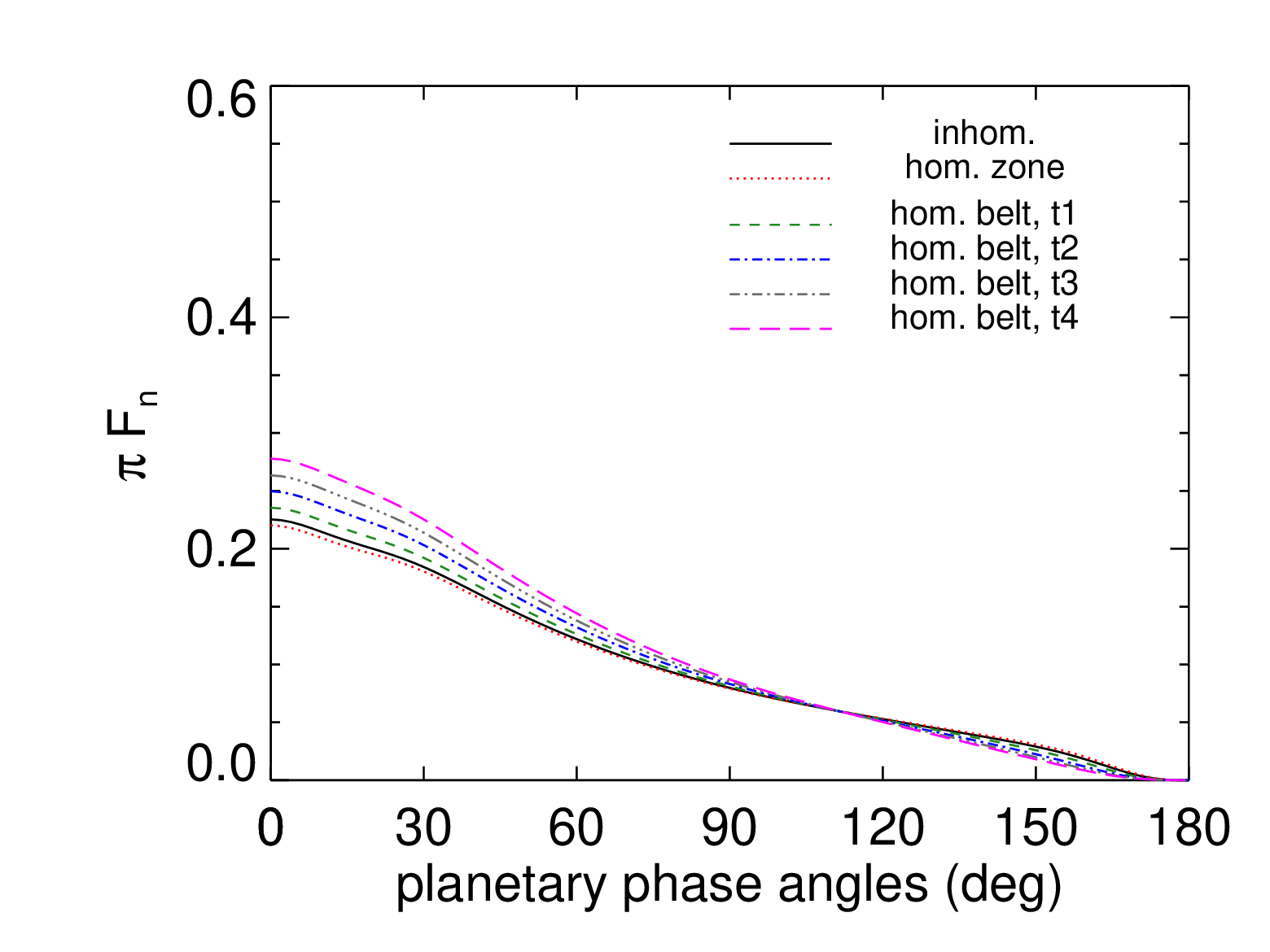}
\hspace{0.8cm}
\centering
\includegraphics[width=85mm]{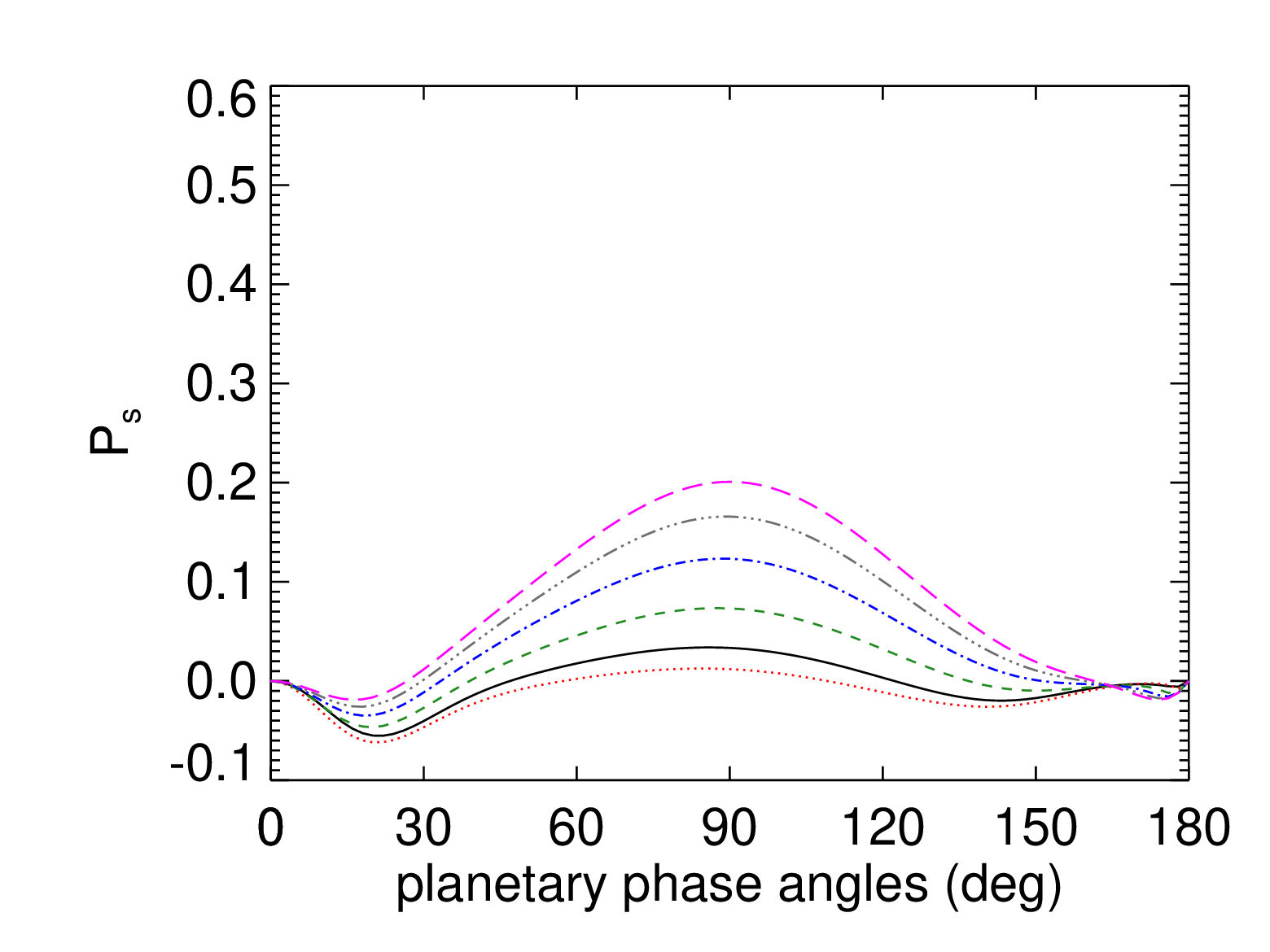}
\caption{Same as in Fig.~\ref{fig:top02_550nm}, except for
         $\lambda=$0.75~$\mu$m.}
\label{fig:top02_750nm}
\end{figure}

\begin{figure}
\centering
\includegraphics[width=85mm]{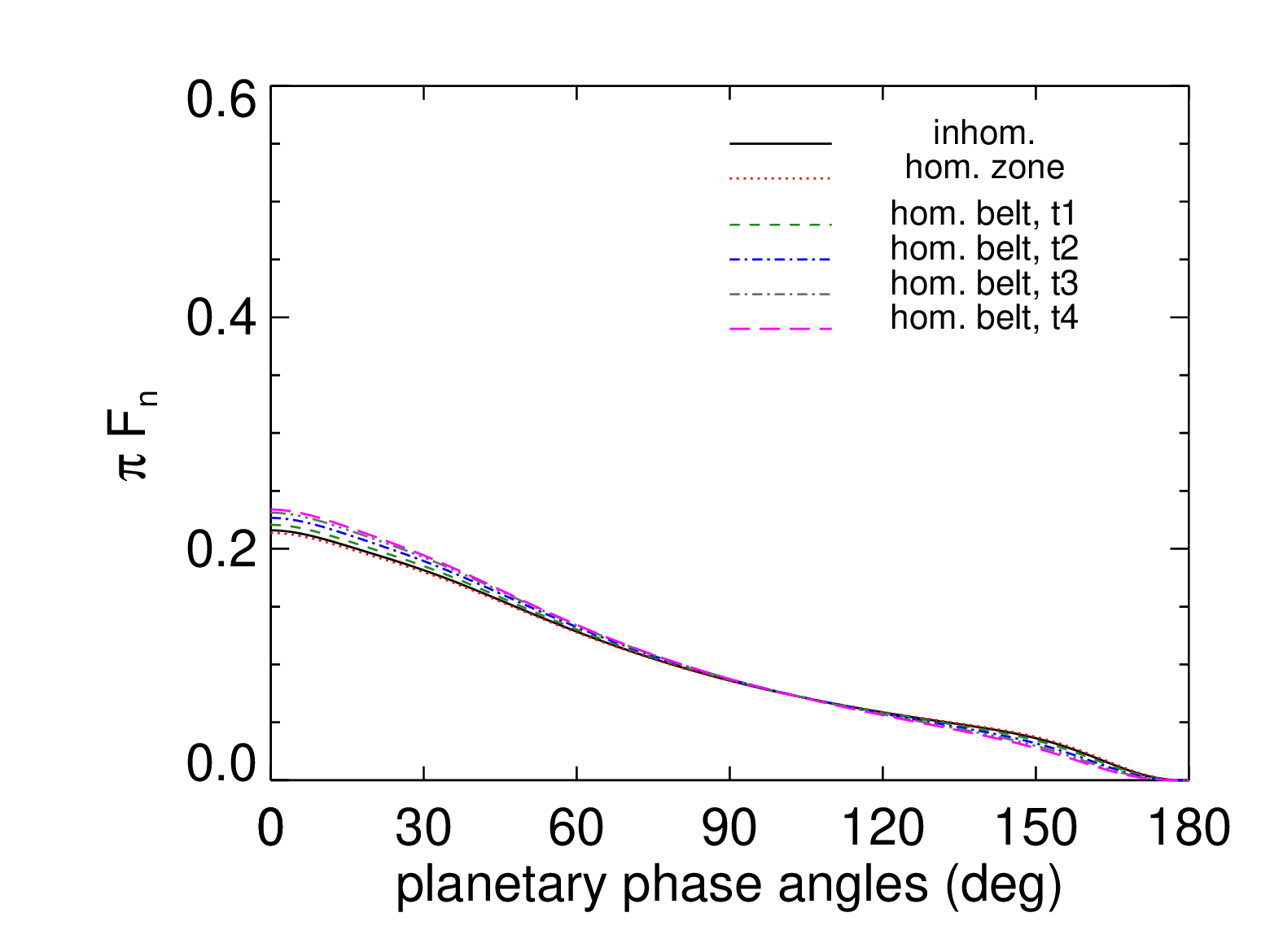}
\hspace{0.8cm}
\centering
\includegraphics[width=85mm]{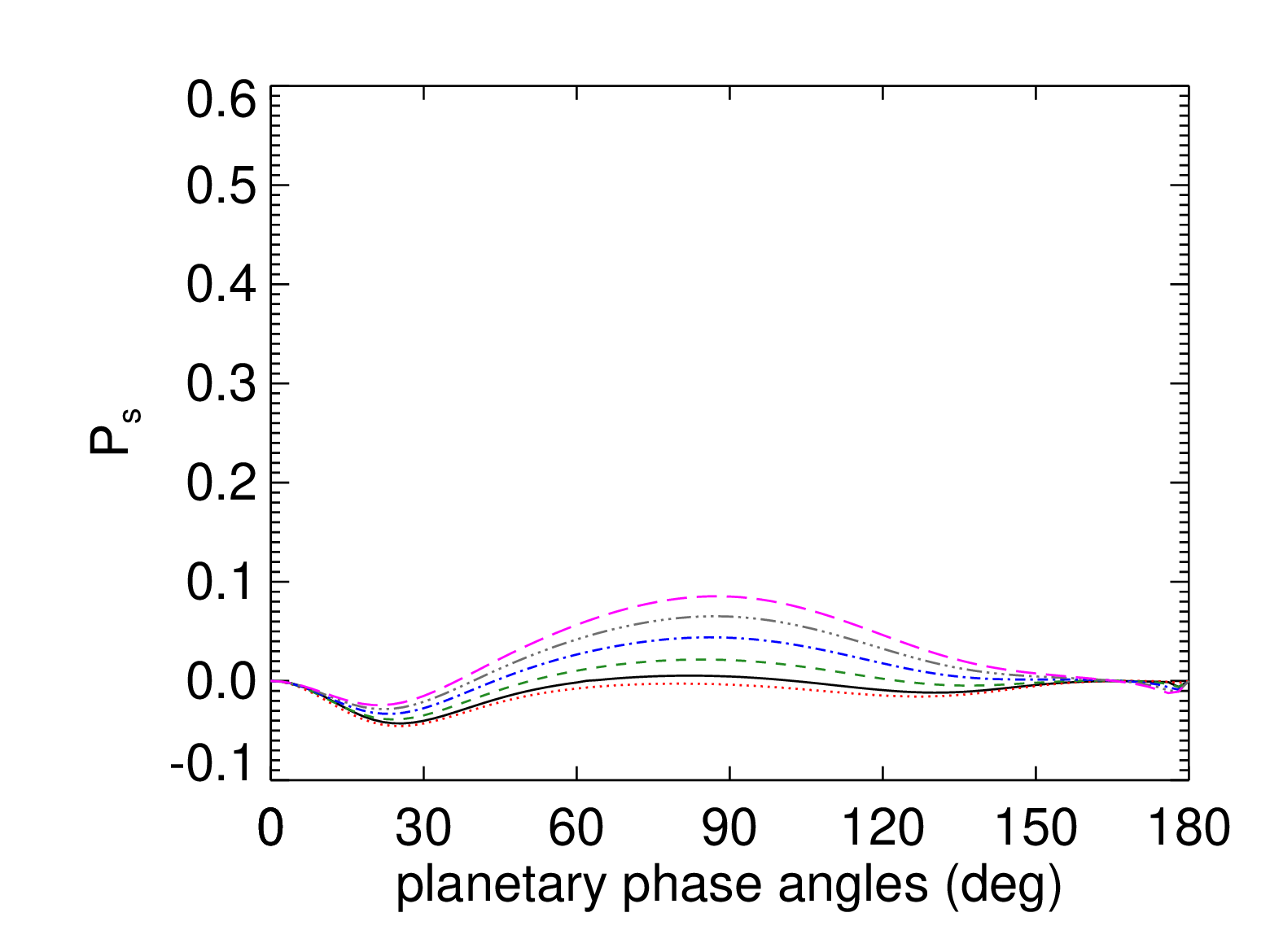}
\caption{Same as in Fig.~\ref{fig:top02_550nm}, except for
         $\lambda=$0.95~$\mu$m.}
\label{fig:top02_950nm}
\end{figure}

The model atmospheres in this section contain only clouds, no hazes.
Figures~\ref{fig:top02_550nm}--\ref{fig:top02_950nm} show the 
flux $\pi F_\mathrm{n}$ and the
degree of linear polarization $P_\mathrm{s}$ as functions of $\alpha$
at $\lambda=0.55~\mu$m (Fig.~\ref{fig:top02_550nm}),
0.75~$\mu$m (Fig.~\ref{fig:top02_750nm}), and
0.95~$\mu$m (Fig.~\ref{fig:top02_950nm}),
for horizontally homogeneous planets with the bottom of the cloud layer
at 1.0~bar, and the top at 0.1, 0.2, 0.3, 0.4, or 0.5~bar.
Also shown in these figures, are $\pi F_\mathrm{n}$ and $P_\mathrm{s}$
for horizontally inhomogeneous model planets each 
with a cloud top pressure of 0.1~bar in the
zones and with cloud top pressures ranging from 0.2~bar (Fig.~\ref{fig:top02_550nm},
~\ref{fig:top02_750nm} and~\ref{fig:top02_950nm})
to 0.5~bar (Fig.~\ref{fig:top03_550nm}) in the belts. The latitudinal borders of the
zones and belts have been described in Sect.~\ref{sect:modelplanets}.


For each model planet and each wavelength,
total flux $\pi F_\mathrm{n}$ at $\alpha=0^\circ$ equals the planet's 
geometric albedo $A_\mathrm{G}$. With increasing wavelength, $A_\mathrm{G}$
decreases slightly, because of the decreasing cloud optical thickness
with $\lambda$, and the decreasing single scattering
phase function in the backscattering direction 
(see Fig.~\ref{fig:ss_ice}).
With increasing $\alpha$, $\pi F_\mathrm{n}$ decreases smoothly for all
model atmospheres. The angular feature around $\alpha= 12^\circ$
for the horizontally homogeneous planets with the highest cloud layers,
can be retraced to the single scattering
phase function (Fig.~\ref{fig:ss_ice}). The strength of the feature
in the planetary phase functions decreases with $\lambda$, just like
that in the single scattering phase functions. The decrease of the
feature with increasing cloud top pressure is due to the increasing 
thickness of the gas layer overlying the clouds.
With increasing $\lambda$, the difference between the total fluxes 
reflected by the model atmospheres decreases, mostly because of the 
decrease of Rayleigh scattering above the clouds with $\lambda$. 

Interestingly, $\pi F_\mathrm{n}$ is insensitive to the cloud top pressure 
around $\alpha=125^\circ$ at $\lambda=0.55$~$\mu$m 
(Fig.~\ref{fig:top02_550nm}). With increasing
$\lambda$, the phase angle where this insensitivity occurs decreases:
from about 110$^\circ$ at $\lambda=0.75~\mu$m 
(Fig.~\ref{fig:top02_750nm}), to about 90$^\circ$ 
at $\lambda=0.95~\mu$m (Fig.~\ref{fig:top02_950nm}). 
Thus precisely across the phase angle range where 
exoplanets are most likely to be directly detected because they
are furthest from their star, reflected fluxes do not give access
to the cloud top altitudes. 


The degree of linear polarization, $P_\mathrm{s}$, shows the typical 
bell-shape around approximately $\alpha=90^\circ$, that is due to 
Rayleigh scattering of light by gas molecules 
(Fig.~\ref{fig:top02_550nm}).
With increasing cloud top altitude, hence decreasing Rayleigh 
scattering optical thickness above the clouds, the
features of the single scattering phase function of the cloud 
particles become more prominent. This is especially obvious at
the longer wavelengths, i.e. at $0.75~\mu$m and 0.95~$\mu$m,
where the Rayleigh scattering optical thickness above the clouds 
is smaller by factors of about $(0.55/0.75)^4$ and $(0.55/0.95)^4$,
respectively (Figs.~\ref{fig:top02_750nm} and~\ref{fig:top02_950nm}).
In particular, the negative polarized feature below
$\alpha=30^\circ$, that is due to light singly scattered by the cloud
particles (see Fig.~\ref{fig:ss_ice}) becomes more prominent.

Figure~\ref{fig:top02_550nm} clearly shows that, unlike the reflected flux,
$P_\mathrm{s}$ is sensitive to cloud top altitudes across 
planetary phase angles that are important for direct detections.
The reason is that $P_\mathrm{s}$ is very sensitive to the Rayleigh
scattering optical thickness above the clouds, as has been known for
a long time from observations of Solar System planets, such as 
the ground-based observations of Venus \citep[][]{hansenhovenier74}, 
and remote-sensing observations of the Earth by instruments such as POLDER
on low--orbit satellites \citep[][]{knibbe00}.
As expected, with increasing $\lambda$, the sensitivity of $P_\mathrm{s}$
to the cloud top altitude decreases (see Figs.~\ref{fig:top02_750nm} 
and~\ref{fig:top02_950nm}).

Figures~\ref{fig:top02_550nm}--\ref{fig:top02_950nm} also show 
$\pi F_\mathrm{n}$ and $P_\mathrm{s}$ of horizontally inhomogeneous planets 
with zones and belts.
In all figures, the cloud top pressure of the zones is 0.1~bar while
that at the top of the belts varies from 0.2~bar 
(Figs.~\ref{fig:top02_550nm}, \ref{fig:top02_750nm}, 
and~\ref{fig:top02_950nm}) to 0.5~bar (Fig.~\ref{fig:top03_550nm}). 
The shapes of the flux and polarization phase functions of these
horizontally inhomogeneous planets are very similar to those
of the horizontally homogeneous planets: one could easily find a 
horizontally homogeneous model planet with a cloud top pressure
between 0.1 and 0.3~bar that would fit the curves pertaining to
the horizontally inhomogeneous planets. The cloud top pressure 
that would provide the best fit would be slightly different when
fitting the flux or the polarization curves. For example, 
fitting the flux reflected by an inhomogeneous planet with cloud 
top pressures in the belts at 0.4~bar (blue dashed--dotted line of 
Fig.~\ref{fig:top03_550nm}) would 
require a homogeneous planet with its cloud top pressure at 0.2~bar,
while fitting the polarization would require a cloud top pressure 
of about 0.18~bar. Such small differences would most likely disappear
in the measurement errors. With increasing $\lambda$, the effects of the
cloud top pressure decrease, in particular in $\pi F_\mathrm{n}$.
Covering a broad spectral region would thus not help in narrowing
the cloud pressures down.

\subsection{Cyclonic spots}
\label{sec:cyclonicspots}

Other spatial features on giant planets in our Solar System 
are (anti--)cyclonic storms that show up as oval--shaped spots. 
Famous examples are Jupiter's Great Red Spot (GRS) that appears
to have been around for several hundreds of years and Neptune's Great 
Dark Spot (GDS) that was discovered in 1989 by Voyager--2, but that seems to
have disappeared \citep{hammel95}. Recent spots on Uranus were 
presented by \citet{hammel09} and \citet{sromovsky12}.
To study the effect of localized spots on reflected flux and polarization
signals of exoplanets,
we use a Jupiter--like model atmosphere with a spot of NH$_3$ ice
clouds extending between 0.75~and 0.13~bar.
The clouds in the spot have an optical thickness of 36 at $\lambda=0.5~\mu$m 
\citep[][]{simonmiller01}.
We model the spot as a square of 26$^\circ$
in longitude by 22$^\circ$ in latitude with an optical thickness of
$\sim$30 at 0.55~$\mu$m.
Additionally, our model planet has zones and belts spatially distributed
across the planet as described before, extending between 0.56~to 0.18~bar 
in the zones, and from 1.0~to 0.5~bar in the belts. The cloud optical thickness 
in the belts is 6.02, and in the zones 21.0 
at $\lambda=0.75~\mu$m.


\begin{figure}
\centering
\includegraphics[width=85mm]{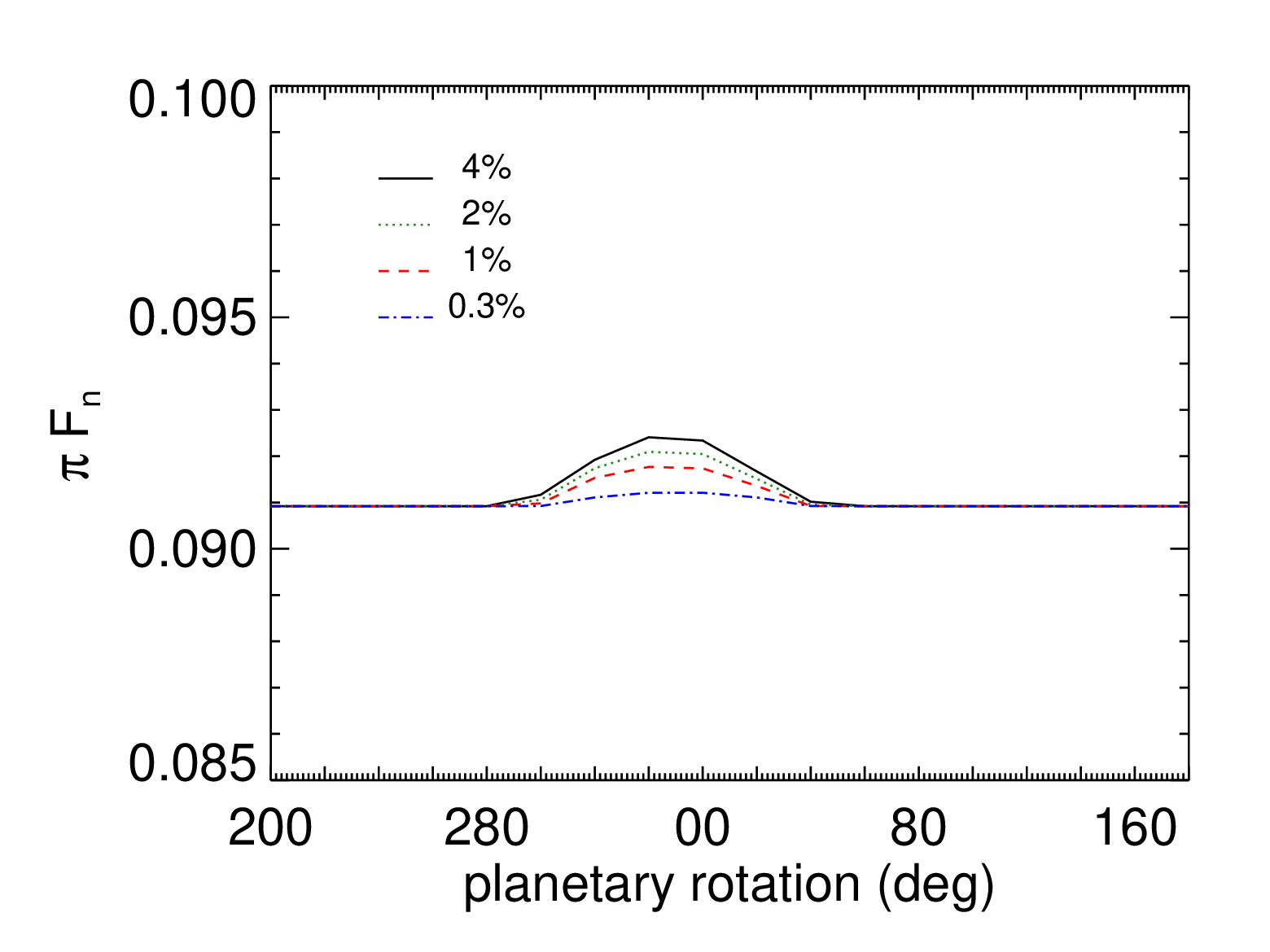}
\hspace{0.8cm}
\centering
\includegraphics[width=85mm]{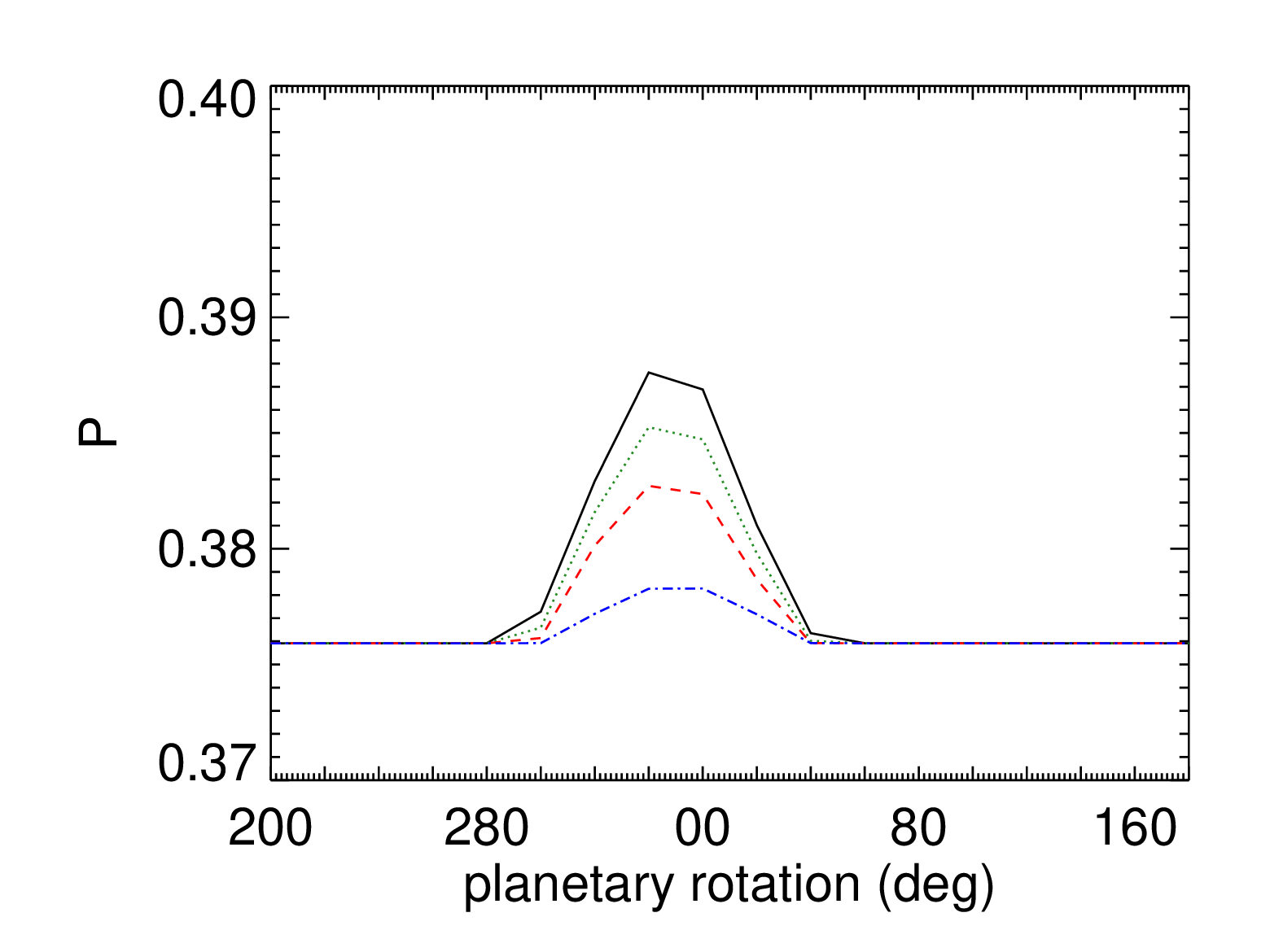}
\caption{$\pi F_\mathrm{n}$ and $P_\mathrm{s}$ as functions of the planet's
         rotation angle (in degrees), for a Jupiter--like model planet 
         with a spot on its equator. Curves are shown for spots covering
         0.3\% (blue, dashed--dotted line), 
         1\% (red, dashed line), 2\% (green, dotted line), or 4\% 
         (black, solid line) of the planetary disk. 
         The planetary phase angle $\alpha$ is
         90$^\circ$ and $\lambda=0.55~\mu$m. Calculations have been done
         at rotation angle steps of 20$^\circ$.}
\label{fig:jup3}
\end{figure}

Figure~\ref{fig:jup3} shows reflected fluxes $\pi F_\mathrm{n}$ and 
degree of polarization $P_\mathrm{s}$ at $\lambda=0.55~\mu$m, 
as functions of the planet's rotation angle.
The planet's phase angle is 90$^\circ$. 
The spot is on the planet's equator
(which coincides with the planetary scattering plane). 
Recall that both Jupiter and Saturn have rotation periods on the order of 
10~hours, while Uranus and Neptune rotate in about 17 and 16 hours, 
respectively. With a 10-hour rotation period and $\alpha=90^\circ$, 
a small spot would cross the illuminated and visible part of the disk
in about 2.5~hours.
Curves are shown for different sizes of the spot: covering at maximum 
0.3~\%, 1\%, 2\%, or 4\% of the planet's disk, respectively.
For comparison: the GRS covers about 6~\% of Jupiter's disk.
Note that the calculations for Fig.~\ref{fig:jup3}
have been done per 20$^\circ$ rotation of the planet, due to computational
restrictions.

As can be seen in Fig.~\ref{fig:jup3}, 
the reflected fluxes $\pi F_\mathrm{n}$ hardly change upon the passage of the 
spot across the illuminated and visible part of the planetary disk:
even for the largest spot located at the equator, the maximum
change in $\pi F_\mathrm{n}$ is a few percent.
In $P_\mathrm{s}$, the transiting spots also leave a change of at
most a few percent (absolute, since $P_\mathrm{s}$ is a relative 
measure itself). For spots located at higher latitudes of the planet,
the effects are even smaller.
With increasing wavelength, the sensitivity of both $\pi F_\mathrm{n}$ and
$P_\mathrm{s}$ to the cloud top altitude decreases. At
longer wavelengths, the effects of a spot would thus be smaller than
shown in Fig.~\ref{fig:jup3}.

\subsection{Polar hazes}
\label{sec:polarhazes}

The poles of Jupiter and Saturn are covered by stratospheric hazes. 
In particular, when seen at phase angles around 90$^\circ$, 
Jupiter's polar hazes yield strongly polarized signals. 
This high polarization can be explained
by haze particles that consist of aggregates of particles that are small
compared to the wavelength, and that polarize the incident sunlight as
Rayleigh scatterers \citep[][]{west91}, with a high degree of polarization
at scattering angles around 90$^\circ$.
We are interested in whether strongly polarized polar hazes will leave
a trace in the disk-integrated polarization signal of a planet.


\begin{figure}
\centering
\includegraphics[width=85mm]{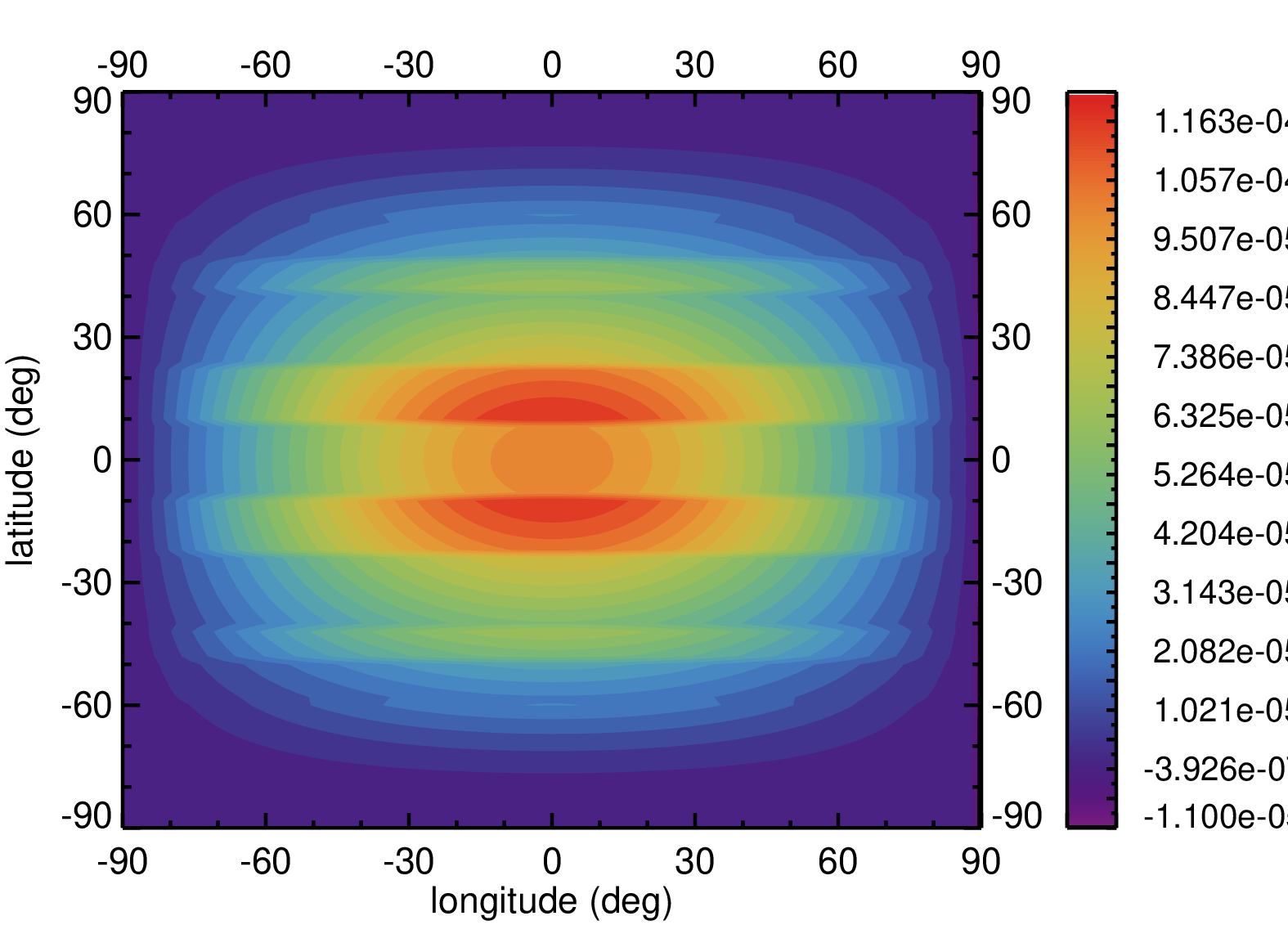}
\hspace{0.8cm}
\centering
\includegraphics[width=85mm]{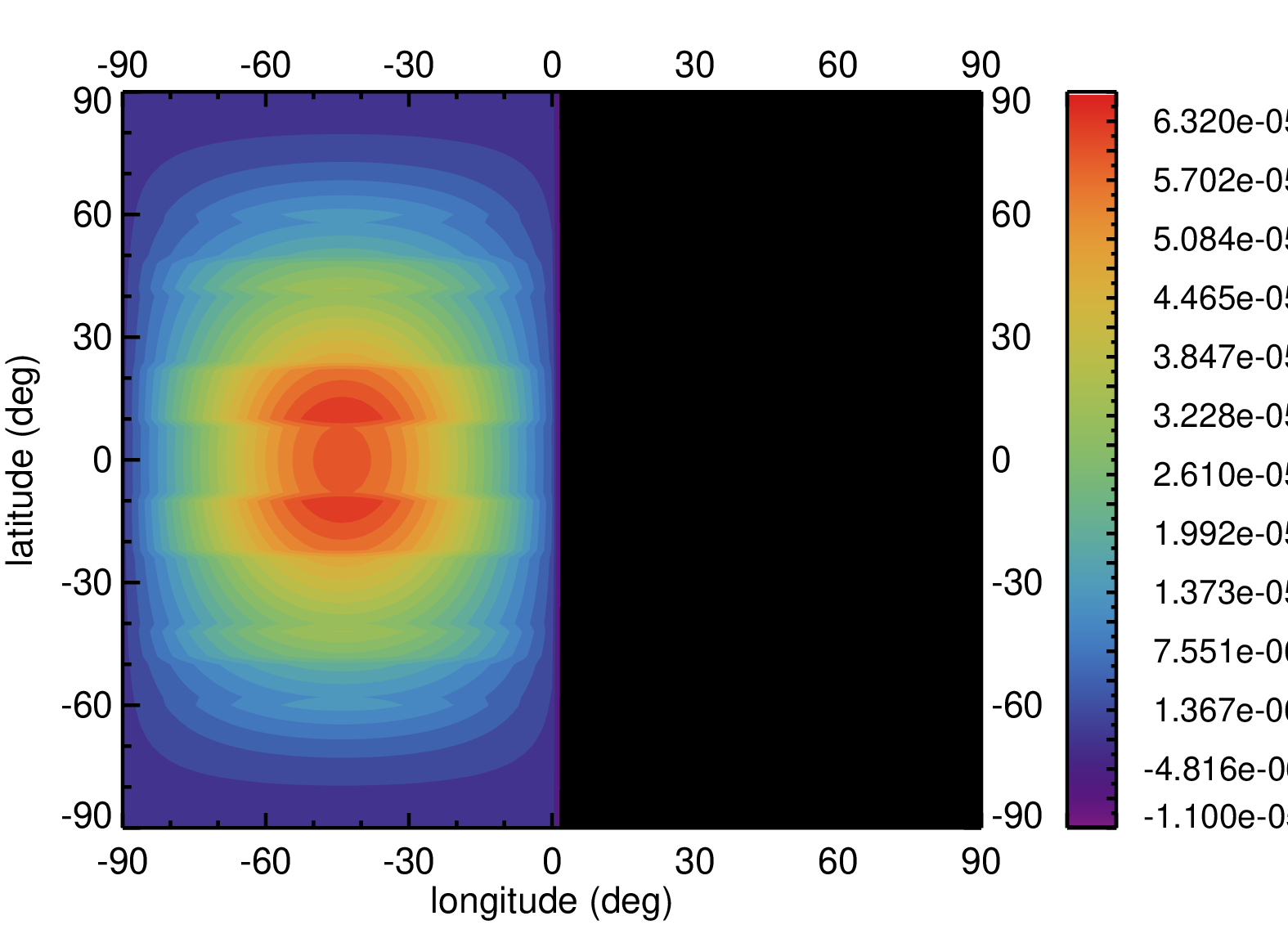}
\caption{Reflected flux $\pi F_\mathrm{n}$ at $\lambda=0.55~\mu$m for
         every pixel on the disk of a
         Jupiter--like model planet containing clouds in zones and belts, 
         and polar stratospheric hazes at latitudes above (below) 60$^\circ$
         (-$60^\circ$). The phase angles are: 0$^\circ$ (top) and 90$^\circ$
         (below).}
\label{fig:flx_a90}
\end{figure}

We use a Jupiter--like model planet with clouds in belts and zones as in
Sect.~\ref{sec:zonesbelts}. The cloud top pressure of the belts is~0.3~bar.
Starting at latitudes of $60^\circ$, the north and south poles of 
each model planet are covered by polar haze particles as described in 
Sect.~\ref{sec:hazeparts}. The optical thickness of the haze is
0.2 at $\lambda=$0.55, 0.75, and 0.95~$\mu$m.

\begin{figure}
\centering
\includegraphics[width=85mm]{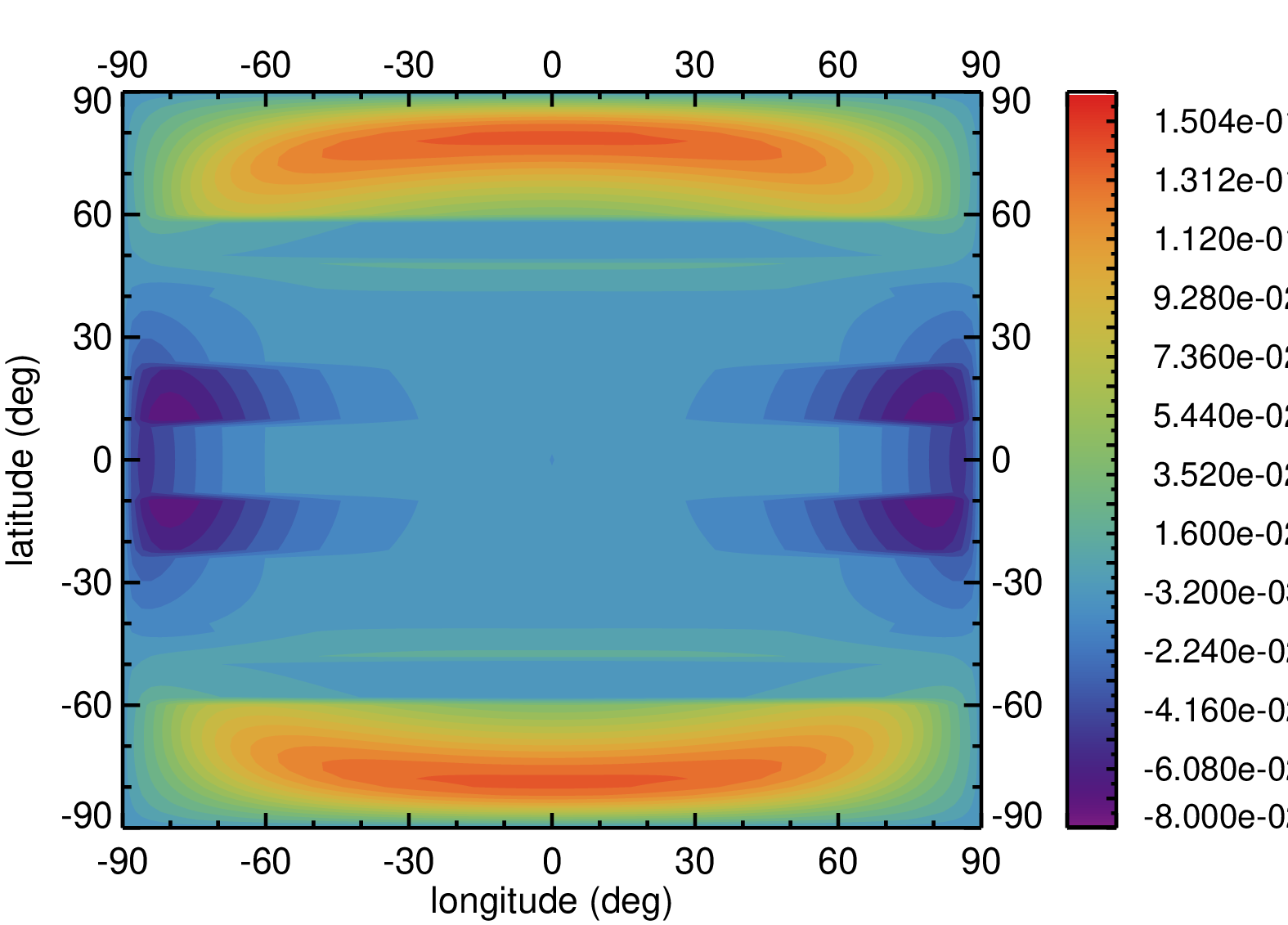}
\hspace{0.8cm}
\centering
\includegraphics[width=85mm]{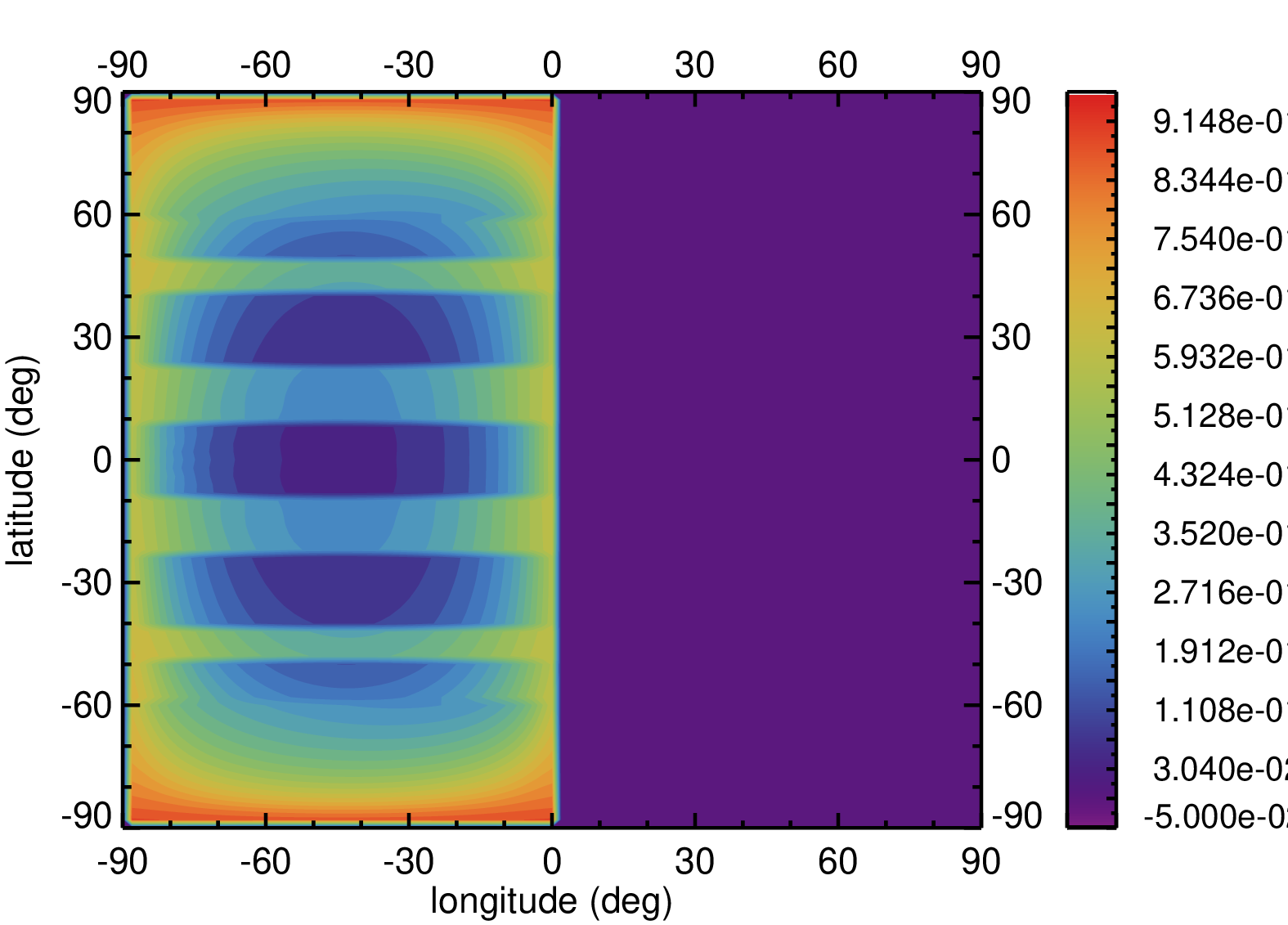}
\caption{Same as in Fig.~\ref{fig:flx_a90} except for $P_\mathrm{s}$.
         Note the different color scales.}
\label{fig:pol_a90}
\end{figure}

\begin{figure}
\centering
\includegraphics[width=85mm]{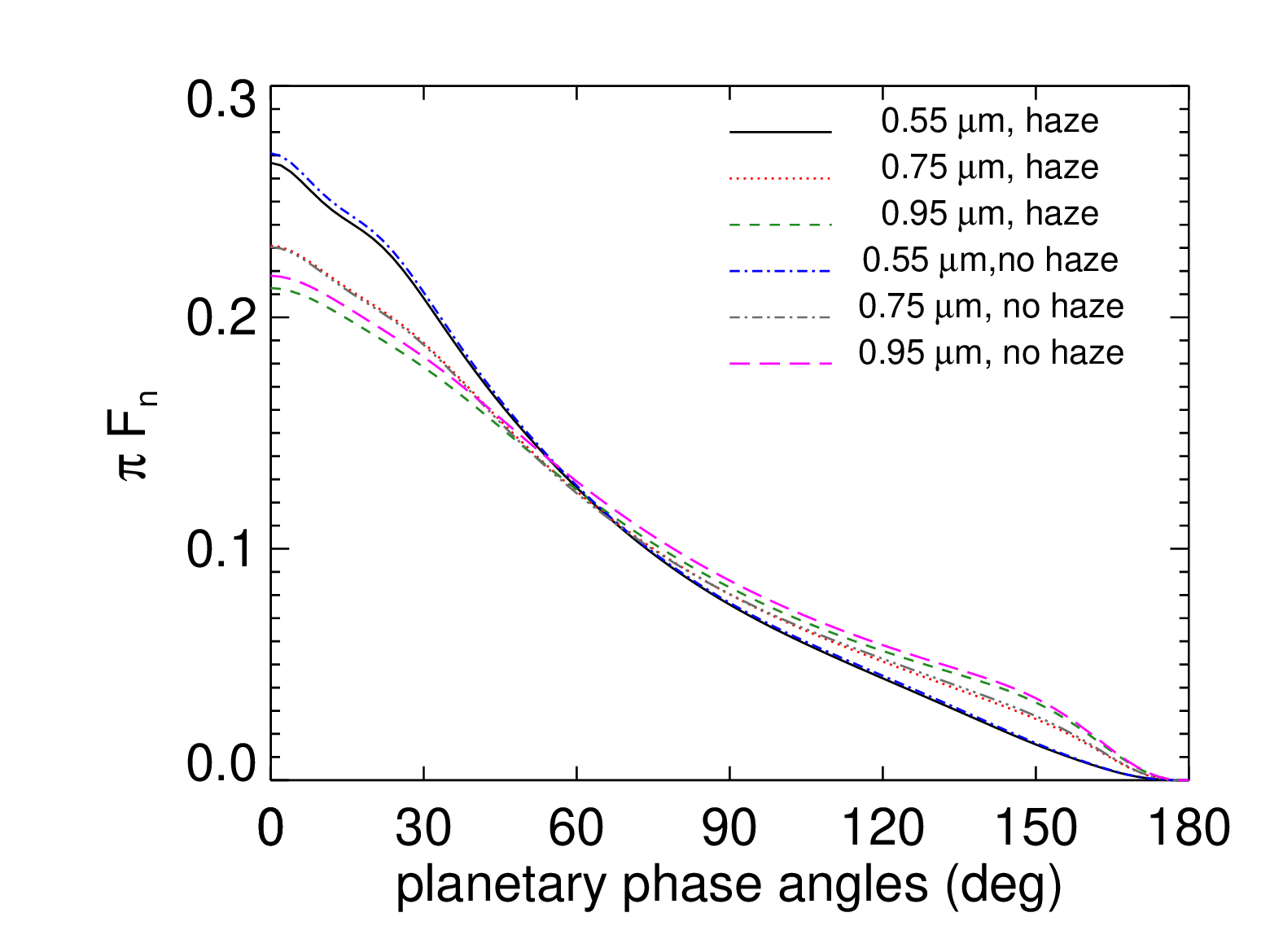}
\hspace{0.8cm}
\centering
\includegraphics[width=85mm]{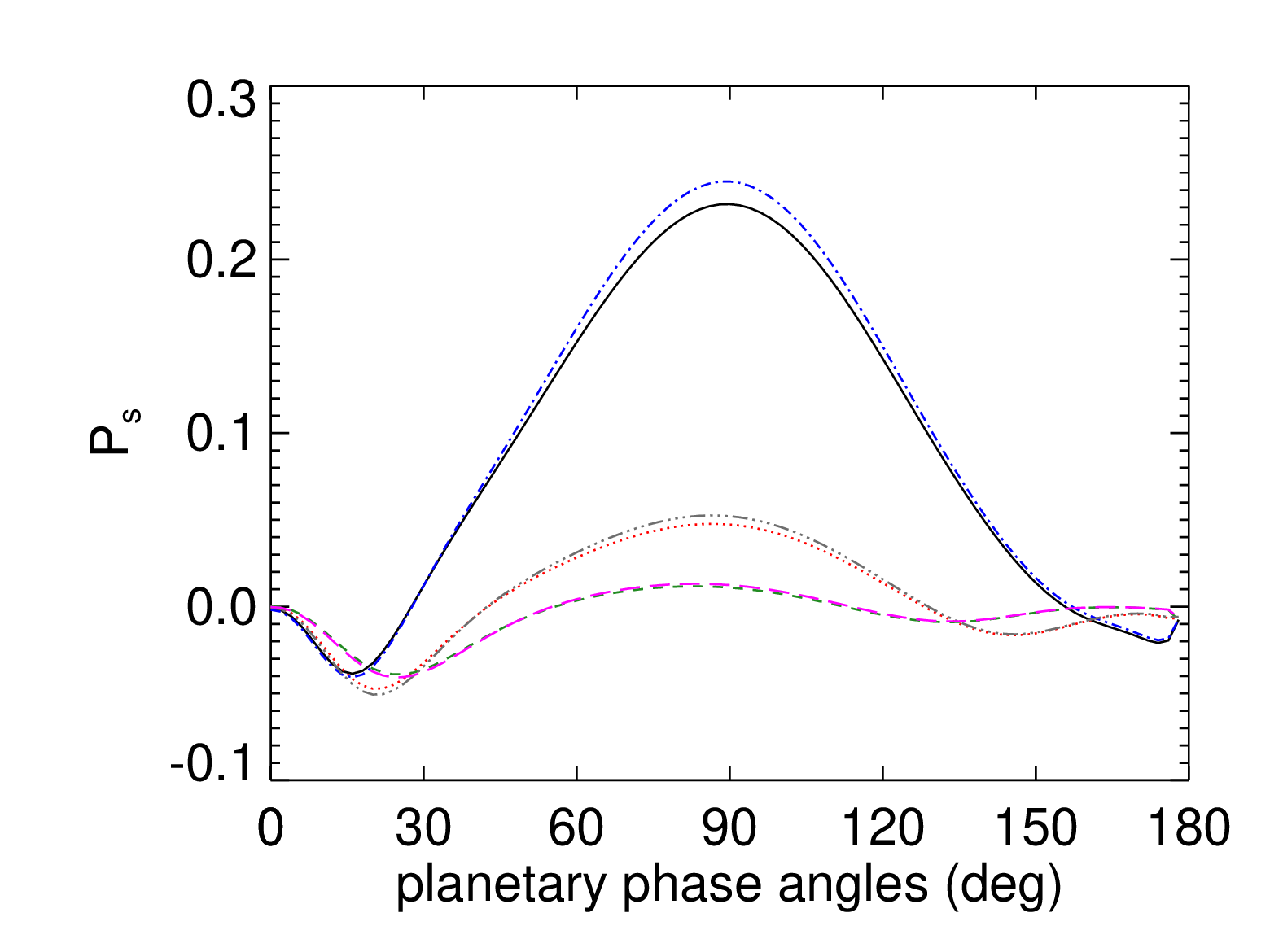}
\caption{Disk--integrated $\pi F_\mathrm{n}$ and $P_\mathrm{s}$ as functions of
         phase angle $\alpha$ for the Jupiter--like model planet with 
         polar hazes as used in 
         Figs.~\ref{fig:flx_a90} and~\ref{fig:pol_a90} at:
         $\lambda= 0.55$~$\mu$m (black, solid lines), 0.75~$\mu$m
         (red, dotted lines), and 0.95~$\mu$m (green, dashed lines).
         Also shown are the lines for model planets with the clouds
         but without the hazes: 
         $\lambda= 0.55$~$\mu$m (blue, dashed--dotted lines), 
         0.75~$\mu$m (grey, dashed--triple--dotted lines), 
         and 0.95~$\mu$m (magenta, long--dashed lines).}
\label{fig:disk_ints}
\end{figure}

In Figs.~\ref{fig:flx_a90} and~\ref{fig:pol_a90}, we show, respectively 
$\pi F_\mathrm{n}$ and $P_S$ at $\lambda=0.55$~$\mu$m of starlight that is
locally reflected by pixels on the visible and illuminated part of 
the planetary disk for $\alpha=0^\circ$ and $90^\circ$.
In Fig.~\ref{fig:disk_ints}, we have plotted the disk--integrated
reflected flux $\pi F_\mathrm{n}$ and degree of polarization $P_\mathrm{s}$
as functions of $\alpha$, for the Jupiter--like model planet with
and without polar hazes, and for $\lambda=0.55$, 0.75,
and 0.95~$\mu$m.

The reflected flux across the planetary disk (Fig.~\ref{fig:flx_a90}) 
shows clear differences between the zones and the belts, 
especially around the center of the planetary disk (for both values of
$\alpha$). The reflected fluxes due to the polar hazes do not stand out 
against those due to the clouds in the belts and the zones.
The reflected flux pattern across the visible and illuminated part
of the planetary disk for $\alpha=90^\circ$
is very similar to that for $\alpha=0^\circ$, with the brightest
regions in the center, and the darkest at the limb and, for $\alpha=90^\circ$,
at the terminator.
Integrated across the planetary disk (Fig.~\ref{fig:disk_ints}), 
there is very little difference between $\pi F_\mathrm{n}$ of the planets
with and without haze. From the disk--integrated reflected fluxes
it would thus be impossible to derive the presence of the polar hazes.
Note that our simulations pertain to broadband fluxes. Observations
across gaseous absorption bands, such as those of methane, could 
provide more information about the presence of hazes, because 
the latter would decrease the depths of the bands 
\citep[see e.g.][]{stamhovenier04}. Of course, in order to use absorption band
depths to derive altitudes of hazes and/or clouds, independent 
information about the mixing ratios of the absorbing gases would be 
essential.

The degree of polarization across the planetary disk (Fig.~\ref{fig:pol_a90}),
also shows clear differences between the zones and the belts, but mostly
near the limb and, for $\alpha=90^\circ$, the terminator of the planet.
For $\alpha=0^\circ$, the observed light has been scattered in the 
backward direction, and the degree of polarization is close to zero 
across the central region of the disk. Towards the limb, the degree
of polarization increases to reach values as high as 0.15 
towards the northern and southern limbs, and then it decreases again. 
The relatively high polarization values are due to significantly polarized
second order scattered light (the singly scattered light contributes virtually
no polarized light), while the low values at the limb are due 
to light that has been singly scattered in the backscattering direction
(cf. Fig.~\ref{fig:ss_ice}).


For $\alpha=90^\circ$, the scattering angle of the singly scattered light
is 90$^\circ$ (cf. Fig.~\ref{fig:ss_ice}). The degree of polarization 
is a few percent at the center of the illuminated and visible part of 
the planetary disk, where the contribution of multiple scattered light
is significant, and increases towards the terminator and the limb.
The higher values (up to 0.90 towards the northern and southern
limbs, and up to 0.50 towards the western limb and terminator) 
are due to singly scattered light. 
We have repeated our calculations for $\lambda= 0.75$ (not shown),
and found that for $\alpha=0^\circ$ the polar hazes produce a 
fractional polarization signal $Q/F$ of $\sim 6$\%, which is 
similar to what was observed in that wavelength range 
by \citet[][]{schmid11}.
The degree of polarization due to the polar hazes will depend 
strongly on the single scattering polarization phase function of 
the particles and hence on the wavelength. 
\citet{kolokolova12} present calculated polarization
phase functions of large fluffy and compact aggregates with application to
polarization observations of comets, and clearly show the increase of 
the maximum degree of polarization around $\alpha=90^\circ$ 
($\Theta=90^\circ$) with increasing $\lambda$ for particles built using
the CCA--method.

While the polar hazes strongly influence the degree of polarization
of local regions on the model planet,
integrated across the planetary disk (Fig.~\ref{fig:disk_ints}),
they have a small ($\lambda=0.55~\mu$m) to
negligible ($\lambda=0.95~\mu$m) influence.
Like with the reflected fluxes (see above), the polar hazes
could have a stronger effect across gaseous absorption bands, 
where the observed reflected starlight originates from higher
altitudes in the planetary atmosphere.


\section{Summary and conclusions}
\label{sect_summary}

In view of upcoming instruments for the direct detections of fluxes
and polarization signals of starlight that is reflected by orbiting exoplanets,
like SPHERE \citep{dohlen08,roelfsema11}
on the VLT, GPI \citep{macintosh08} on the Gemini North telescope, and,
further in the future, EPICS \citep[][]{kasper10} on the E-ELT,
we have presented numerically calculated disk--integrated, 
broadband flux and polarization signals 
of horizontally inhomogeneous gaseous exoplanets in order to investigate whether
or not spatial inhomogeneities such as those found on gas giants in the 
Solar System could be identified on gaseous exoplanets.
The spatial inhomogeneities that we have modeled are: cloud top altitudes
belts and zones due to varying cloud top altitudes (Sect.~\ref{sec:zonesbelts}),
cyclonic spots (Sect.~\ref{sec:cyclonicspots}), 
and polar hazes (Sect.~\ref{sec:polarhazes}).

We have calculated the total flux and polarization signals of the exoplanets
due to the mentioned
spatial inhomogeneities using an adding--doubling radiative transfer 
code that fully includes all orders of scattering and linear and circular
polarization. Here, we ignore the circular polarization because it is 
very small and neglecting it does not introduce significant errors in the
calculated total and linearly polarized fluxes \citep[][]{stam05}. 
Circular polarization will also not be measured by the above mentioned instruments.
We only consider exoplanets in wide orbits, such that they can be
spatially resolved from their parent star, and ignore
diffracted light from the parent star and starlight that is scattered
by e.g. exozodiacal dust in the planetary system (we thus assume that
the observations can be corrected for these two types of background signals).

A version of the adding--doubling radiative transfer code that handles 
vertically inhomogeneous,
but horizontally homogeneous exoplanets has been described by 
\citet{stamhovenier04}.
In our version of the code \citep[see][]{karalidi12b,karalidi12c}, we divide a
horizontally inhomogeneous planet into horizontally homogeneous pixels. For each
pixel, we calculate the reflected total and polarized fluxes and add
up these local fluxes (including rotations between local and planetary
reference planes for the polarized fluxes) to obtain the disk--integrated 
total and polarized fluxes at a given planetary phase angle.
We compare signals calculated for horizontally inhomogeneous planets
with those of horizontally homogeneous planets to investigate the
appearance of the spatial inhomogeneities in the disk--integrated
signals.

Cyclonic spots on a planet will rotate in and out of the view of the observer
depending on their daily rotation periods
(because we consider exoplanets in wide orbits, they are unlikely to be 
tidally locked to their star), and will thus give rise to time--varying
total and polarized fluxes.
Our model exoplanets with zones and belts, and those with the polar hazes
are symmetric with respect to their axis of rotation. The total 
flux and polarization signals of these planets will thus depend on the
planetary phase angle, but not on daily rotations of a planet.
For these model exoplanets, the presence of horizontal inhomogeneities
could presumably be derived from different effects the inhomogeneities
have on the total flux and on the degree of linear polarization of 
the reflected starlight.

Our Jupiter--like model exoplanets have thick cloud layers consisting of
spherical ammonia ice particles. In Sect.~\ref{sec:zonesbelts}, we model 
belts and zones by choosing 
different pressure levels for the tops of the clouds, while the base of the clouds
is fixed at 1~bar. The pressure at the top of the zonal clouds is fixed at 
0.1~bar, while the pressure at the top of the belts is varied between 0.2~bar
and 0.5~bar. Comparing the total and polarized fluxes of the horizontally
inhomogeneous planets with those of horizontally homogeneous cloudy planets,
it is clear that both the reflected total fluxes and the polarization are
sensitive to the cloud top altitude. The polarization appears to be more
useful to derive cloud top altitudes than the reflected flux, because the
polarization is most sensitive to the cloud top altitude at planetary 
phase angles around 90$^\circ$, which are favorable for direct imaging
of exoplanets. At these phase angles, the sensitivity of the total flux
to the cloud top altitude is small, and it decreases with increasing wavelength,
because of the decreasing Rayleigh scattering optical thickness above the
clouds. 

The shape of the polarization phase function for a planet with belts and 
zones is very similar to that of a horizontally homogeneous planet with
a cloud top pressure between that of the zones and the belts (the precise 
value depends on the latitudes covered by the belts and zones and by the
cloud top pressures across the belts and the zones). Combining polarization
phase functions at different broadband wavelengths will not help to 
reveal the presence of horizontal inhomogeneities. Combining total 
flux phase functions with polarization phase functions will also fail 
to reveal the presence of inhomogeneities because of the lack of 
sensitivity of the total flux phase functions to the cloud top
pressures, especially at the phase angles that are favorable for 
polarimetry (by lack of independent information about
the planet radius, the total flux phase functions will also have 
error bars that preclude deriving meaningful cloud top pressures from them). 

We model a cyclonic spot as a localized cloudy region with a lower cloud
top pressure than the surrounding clouds: in the spot, the cloud top 
pressure is 0.13~bar, while outside the spot, the cloud top pressure
is 0.5~bar at the edges and 0.18~bar at the central latitudes. 
Our results for model planets with a cyclonic spot on the planet's
equator that covers about 4\% of the planetary disk show that the 
change in the reflected total flux as the spot moves in and out of view
of the observer is less than a percent at a planetary phase angle of 90$^\circ$.
The change in the degree of polarization is a few percent (in absolute
sense). The temporal changes in the total flux and the degree of
polarization decrease if the spot is located further away from the 
planetary scattering plane. If a planet has more than one spot, a 
temporal analysis of flux and polarization time series might reveal 
the locations and sizes of the most prominent spots.

Polar hazes that very likely consist of aggregates of small monomers cover the
poles of Jupiter and Saturn, yielding a locally relatively high degree of 
polarization \citep[][]{west91}. We modeled polar hazes on Jupiter--like
planets using fluffy aggregates covering latitudes north and southwards
of, respectively, 60$^\circ$ and -60$^\circ$. In the locally reflected
total fluxes (i.e. spatially resolved across the planet), 
the polar hazes do not leave a significant trace, and, not surprisingly, 
neither do they in the disk-integrated total fluxes.
In the degree of polarization of the locally reflected light, the polar
hazes do show up. At a phase angle of 0$^\circ$, light that has been
scattered twice by the haze particles has a relatively high degree of 
polarization (compared to the light that has been singly scattered
in the backward direction by the gas molecules or the haze particles, 
and that is virtually unpolarized). At a phase angle of 90$^\circ$,
light that has been singly scattered by the haze particles has a 
high degree of polarization, because the monomers that form the 
aggregated haze particles scatter like Rayleigh scatterers. 

When integrated over the planetary disk, however, the polar hazes
change the degree of linear polarization by only a few percent
when compared to a planet without polar hazes at $\lambda=0.55$~$\mu$m,
and less at longer wavelengths. The shape of the polarization phase
function of the planet with polar hazes is similar to that without
hazes.

Finally, we note that the vast majority of giant planets 
discovered so far are very hot \citep[see e.g.][]{knutson12, demooij13} and 
their atmospheric chemistry and temperature--pressure profiles vary considerably 
from those of the giant planets of our solar system 
\citep[see e.g.][and references there in]{moses11, huitson12, madhusudhan12b}. 
While these close-in planets orbit too close to their parent star 
to be directly detectable and are thus out of the scope of this paper, they 
have taught us that giant planets can exhibit a large variety of properties. 
Even the exoplanets that orbit a few AU from their parent star, and that 
could thus be directly detectable, might be shown in the near future to 
vary considerably in their properties.







\end{document}